\documentclass[prd,nofootinbib,floats,superscriptaddress,eqsecnum,tightenlines,11pt]{revtex4}

\usepackage{hyperref}
\usepackage{graphicx}
\usepackage{amsmath,amssymb,amsfonts,amsthm,latexsym,stmaryrd}
\usepackage{marginnote}
\usepackage{color}
\usepackage{ulem}

\def\beq{\begin{equation}}
\def\eeq{\end{equation}}
\newcommand{\bea}{\begin{eqnarray}}
\newcommand{\eea}{\end{eqnarray}}
\def\bi{\begin{itemize}}
\def\ei{\end{itemize}}
\def\ba{\begin{array}}
\def\ea{\end{array}}
\def\bfig{\begin{figure}}
\def\efig{\end{figure}}

\begin{document}

\title{Reconsidering the Ostrogradsky theorem:  \\
Higher-derivatives Lagrangians, Ghosts and  Degeneracy}

\author{Alexander Ganz}
\affiliation{Dipartimento di Fisica e Astronomia ``G. Galilei'',Universita degli Studi di Padova, via Marzolo 8, I-35131 Padova, Italy}
\affiliation{INFN, Sezione di Padova,via Marzolo 8, I-35131 Padova, Italy}
\author{Karim  Noui}
\affiliation{Institut Denis Poisson, CNRS, Universit\'e de Tours, 37200 Tours, France}
\affiliation{Astroparticule et Cosmologie (APC), 
CNRS, Universit\'e de Paris, F-75013 Paris, France}
\affiliation{Laboratoire de Physique de l'Ecole Normale Sup\'erieure, CNRS, PSL Research University and Sorbonne Universit\'es, 24 rue Lhomond, 75005 Paris, France}

\date{\today}

\begin{abstract}
We review the fate of the Ostrogradsky ghost in higher-order theories. 
We start by recalling the original Ostrogradsky theorem and illustrate, in the context of classical mechanics, how higher-derivatives
Lagrangians lead to unbounded Hamiltonians and then lead to (classical and quantum) instabilities. Then, we extend the Ostrogradsky theorem to higher-derivatives theories  of several dynamical variables and  show the possibility to evade 
the Ostrogradsky instability when the Lagrangian is ``degenerate'', still in the context of classical mechanics. 
In particular, we explain why  
higher-derivatives  Lagrangians and/or higher-derivatives Euler-Lagrange equations do not necessarily lead to the propagation of an Ostrogradsky 
ghost. We also study some quantum aspects and illustrate how the Ostrogradsky instability shows up at the quantum level.
Finally, we generalize our analysis to the case of higher order covariant theories where, as the Hamiltonian is vanishing and thus bounded,
the question of Ostrogradsky instabilities is subtler.

\end{abstract}

\maketitle

\newpage

\tableofcontents

\newpage

\section{Introduction}

In classical mechanics, a Lagrangian with higher (than two) time derivatives of the dynamical variables, that cannot be removed by integrations by part, is often considered as pathological. 
It is commonly believed to lead to the propagation of the celebrated Ostrogradsky ghost in the theory, an unstable degree of freedom. 
This is indeed true in the context of classical mechanics where the Lagrangian depends on only one dynamical variable $\phi(t)$ and its derivatives:
the presence of  higher  time derivatives (which do not reduce to total derivatives) leads systematically to an instability. More precisely, the well-known Ostrogradsky theorem \cite{Ostrogradsky:1850fid} states that its Hamiltonian is not bounded neither from above nor from below. The fact that the Hamiltonian is unbounded can be shown to be related to the existence of
new degrees of freedom in the theory, the Ostrogradsky ghosts,  which makes the theory fundamentally unsafe,  as it is well illustrated in \cite{Woodard:2015zca}. It is easy to understand the presence of these new degrees of freedom from the analysis of the equations of motion which are higher order: a Lagrangian with second derivatives of $\phi$, for instance, leads to a  fourth order equation of motion for $\phi$, which needs
obviously four initial conditions to be integrated, instead of the ordinary two initial conditions of a usual second order equation of motion. Hence, the theory propagates an additional degree of freedom.

Notice that the original Ostrogradsky theorem has been established for Lagrangians which depend on an unique dynamical variable $\phi$
in the context of classical mechanics, where $\phi$ is not a field but a function of time $t$ only, whereas it has been shown that the Ostrogradsky 
ghosts could be avoided for higher order field theories and/or theories with multiple fields.  Indeed, in both cases,  the fact that the 
Lagrangian involves non-trivial higher derivatives of the dynamical fields and the fact that the Euler-Lagrange equations are higher order 
do not necessarily lead to the propagation of Ostrogradsky ghosts.

The celebrated theories of Galileons \cite{Nicolis:2008in,Deffayet:2009wt,Deffayet:2009mn} which have been inspired by the Dvali-Gabadadze-Porrati model \cite{Dvali:2000hr}, are certainly the most popular examples of field theories where the presence of higher derivatives in the Lagrangian does not lead necessarily to Ostrogradsky instabilities. The Galileon is a scalar field, evolving on a fixed background, whose
dynamics is governed by a Lagrangian with second order (space-time) derivatives of the scalar field. 
Even though the second derivatives cannot be removed from the Lagrangian by integrations by part, the equation of motion is still second order, which insures the absence of any  Ostrogradsky ghosts in the theory.  In fact, the Lagrangian has been designed so that the 
corresponding equation of motion remains second order. The generalization of such a construction to  the case where the background becomes itself dynamical has lead to the rediscovery of the famous Horndeski theories \cite{Horndeski:1974wa,Deffayet:2011gz,Kobayashi:2011nu,Deffayet:2013lga} which provide the most general higher-order scalar-tensor theories whose equations of motion remain second order.  
Hence, in these theories, 
requiring second order equations of motion is the key ingredient to evade the Ostrogradsky instability. 
Until a few years ago, it was then commonly believed that only theories that yield second-order Euler-Lagrange equations, for both the scalar field and the metric, were free of this dangerous extra degree of freedom. 
However, it was recently realized  that  this  criterion  is  too  restrictive  and  that  there  actually  exists  a  much  larger  class  of  safe theories, including theories that lead to higher-order Euler-Lagrange equations \cite{Gleyzes:2014dya,Zumalacarregui:2013pma,Lin:2014jga}.  What characterizes these theories is not the order of their Euler-Lagrange equations  but the degeneracy of their Lagrangian, which guarantees the absence of Ostrogradsky instabilities.
In the context of scalar-tensor theories, they have been dubbed Degenerate Higher-Order Scalar-Tensor (DHOST) theories \cite{Langlois:2015cwa,Langlois:2015skt,Achour:2016rkg,Motohashi:2016ftl,BenAchour:2016fzp,Crisostomi:2016tcp,Crisostomi:2016czh,Deffayet:2015qwa,Langlois:2018dxi,Langlois:2020xbc}
which have been at the center of a great activity these last years.
Interestingly,  the issue of Ostrogradsky ghosts can also be addressed in  infinite higher derivative theories \cite{Kolar:2020ezu,Biswas:2011ar,Biswas:2005qr} which are somehow a generalization of higher derivative theories.

To illustrate the notion of degeneracy in the context of higher-order theories, it is simpler to consider examples from classical mechanics. 
Let us start with a Lagrangian of the form $L(\ddot\phi,\dot\phi,\phi,\dot x,x)$
which couples two variables, a ``regular'' one $x$ (with at
most first order derivatives in $L$) and an ``exotic'' one $\phi$ (with second derivatives in  $L$). 
Without any restrictions on the Lagrangian, such a theory admits  three degrees of freedom and needs six initial conditions for the equations
of motion to be fully integrated. One of these three degrees of freedom is an Ostrogradsky ghost.
However, in some situations, the presence of 
$\ddot\phi$ in the Lagrangian that cannot be removed with an integration by part, does not lead necessarily 
to the propagation of an Ostrogradsky ghost. This is the case, for instance, when the Lagrangian is of the form
$L(\dot x + \alpha \ddot\phi,\dot\phi,\phi,x)$, $\alpha$ being a constant,  where we see that the second derivatives can be totally
absorbed into the redefinition of variable $x \rightarrow X \equiv x + \alpha \dot \phi$. This is an example of degenerate Lagrangian but, in general,
a Lagrangian of the form $L(\ddot\phi,\dot\phi,\phi,\dot x,x)$ is said degenerate when there exists a partial differential equation of the form
\bea
\label{constraintF}
F\left(\frac{\partial L}{\partial \ddot \phi} \, , \, \frac{\partial L}{\partial \dot x} \right) = 0 \, ,
\eea
where $F(z_1,z_2)$ is a non-trivial function of the two variables $z_1$ and $z_2$. Such a condition insures that the theory admits, at most, 
two degrees of freedom and then (if in addition $\partial F/\partial z_1 \neq 0$) enables us to get rid of the Ostrogradsky ghost.
Thus, the degeneracy allows to evade the Ostrogradsky instability.  From the Hamiltonian point of view, the condition \eqref{constraintF} is a primary constraint in the phase space of the theory and can be generalized to theories with an arbitrary number of dynamical variables \cite{Motohashi:2016ftl}, to multi-scalar field theories \cite{Crisostomi:2017aim} and to scalar-tensor theories \cite{Langlois:2015cwa}. 
As a consequence, higher-order derivatives in the Lagrangian and higher-order equations of motion is not necessarily a sign
of the presence of an Ostrogradsky ghost, but non-degenerate theories do have Ostrogradsky instabilities.
Very recently, the relation between  the (no-)ghost instability and the degeneracy of a higher-order Lagrangian has been developed in
details in \cite{Aoki:2020gfv}. 
\medskip

The aim of this article is two-fold: reviewing known but subtle (and quite reccent) aspects of higher-order Lagrangians  
and presenting new original results on this topic.   Section \eqref{Ostrotheo} is mainly a review on the Ostrogradsky theorem applied
to classical mechanics where we recall, in particular, the well-known fact that the Ostrogradsky ghost manifests itself through the unboundedness of the Hamiltonian (by the famous linear instability). We provide a complete proof of this theorem. In section \eqref{sec:Degeneracy} we first recall how the degeneracy enables us to evade the Ostrogradsky instabilities, still in the context of classical
mechanics. We consider the general case  of theories with multiple dynamical variables whose
Lagrangian involves non-trivial higher-order derivatives  following \cite{Motohashi:2016ftl}. We also provide a new result and  we derive a general implicit solution of the degeneracy constraints confirming that the higher order Euler Lagrange equations  can be reduced to a system of second order equation of motions. In section \eqref{quantum}, we review quantum aspects of the Ostrogradsky 
instability and we focus mainly on the quantization of the Pais-Uhlenbeck model \cite{Pais:1950za} which provides the simplest free Lagrangian
with higher-derivatives classically stable. We follow the more recent analysis of \cite{Smilga:2017arl,Smilga:2004cy,Mannheim:2004qz}.  We extend the analysis by considering a coupling to a "regular" variable and to show how the coupling triggers the quantum instability. 
Finally, we discuss in section \eqref{sec:covtheories}, the fate of the Ostrogradsky instability
in covariant theories where the Hamiltonian vanishes on-shell, thus it cannot be unbounded from below, and then the question
of Ostrogradsky instabilities become rather subtle. This section is original and contains new results. 
We conclude in section \eqref{conclusion} with a summary and a discussion. 

\section{Ostrogradsky theorem: the linear instability}
\label{Ostrotheo}
In this section,  we review the well-known Ostrogradsky theorem in the context of classical mechanics. We start, in a first subsection, 
by illustrating the theorem 
with the simplest example of  a free (quadratic) Lagrangian for a single field with second derivatives. Then, in a second subsection,
we extend the analysis to the case of a general Lagrangian $L(\ddot\phi,\dot\phi,\phi)$ where we show that, without further restrictions, an Ostrogradsky ghost propagates.

\subsection{The standard model of Ostrogradsky}
The standard (and simplest) model to illustrate the Ostrogradsky instability  describes a point particle system whose 
dynamical variable $\phi(t)$ is governed by a (free) quadratic action of the form
\bea
\label{standardmodel}
S_0[\phi] \; = \; \frac{1}{2} \int dt \, \left( \ddot{\phi}^2 + \alpha \, \dot \phi^2 + \beta \, \phi^2 \right) \, , 
\eea
where $\alpha$ and $\beta$ are constant. They have respectively the dimensions of $\omega^2$ and $\omega^4$ where $\omega$
is a frequency. The equation of motion (see the general form of the Euler-Lagrange equations  for an arbitrary Lagrangian in \eqref{eq:Lagrange_Equation_general}) is simply given by,
\bea
\ddddot{\phi} - \alpha \, \ddot{\phi} + \beta \, \phi \; = \; 0 \, .
\eea
Thus, it is fourth order and needs four initial conditions to be solved which means that two degrees of freedom are propagating in
this theory. One of these two degrees of freedom is the famous Ostrogradsky ghost but, at this stage, we do not really see what is 
wrong with such an extra mode. One may wonder whether there is a difference with a system of two free (coupled or uncoupled) 
oscillators, for instance.

There are different ways to understand the nature of the problem with the Ostrogradsky ghost. All of them rely on a disentanglement between 
the two degrees of freedom propagating in the theory \eqref{standardmodel}. A simple method to proceed consists in considering, 
instead of \eqref{standardmodel}, the equivalent action with no higher derivatives,
\bea
S_{1}[\phi,\psi] \; = \; \frac{1}{2} \int dt \, \left( 
-2 \dot \phi \dot \psi - \psi^2 + \alpha \dot \phi^2 + \beta \phi^2 
\right) \, ,
\eea
which describes two coupled particles $\phi(t)$ and $\psi(t)$. If one integrates out the variable $\psi(t)$ using the equation of motion (for $\psi$) $\psi=\ddot\phi$, one obtains immediately \eqref{standardmodel} up to a boundary term that we neglect. The two actions are therefore classically equivalent, as we claimed above, but this new formulation enables us to see immediately that the theory admits a ghost  when we diagonalize the kinetic term as follows,
\bea
-2 \dot \phi \dot \psi  + \alpha \dot \phi^2 \, = \,   \alpha (\dot\phi - \frac{1}{\alpha} \dot\psi)^2 - \frac{1}{\alpha} \dot\psi^2  \, .
\eea
Hence, the two kinetic terms associated to the two degrees of freedom come with opposite signs. Here, we implicitly assumed that $\alpha \neq 0$ but the case $\alpha=0$ can be treated in a very similar way using the trivial identity 
\bea
\dot \phi \dot \psi=\frac{1}{4} \left[(\dot \phi + \dot \psi)^2-(\dot\phi-\dot\psi)^2\right] \, .
\eea
As a consequence, when $\alpha >0$, $\psi$ is a ghost, and when $\alpha<0$, the combination $\phi- \psi/\alpha$ is a ghost. This is a simple illustration of the celebrated Ostrogradsky ghost and we easily see that the energy of the system is not bounded neither from below nor from above.

This analysis is very instructive to illustrate the nature of the Ostrogradsky instability, but it cannot be, generically, adapted to non-linear theories where the action does not reduce to a functional  quadratic in the variable $\phi$. In the general case, when the Lagrangian is an 
arbitrary function $L(\ddot \phi,\dot\phi,\phi)$, one needs a Hamiltonian analysis to prove the existence of the instability.  Before studying
in details the system governed by a general action $L(\ddot \phi,\dot\phi,\phi)$, let us illustrate the Hamiltonian method with the linear model \eqref{standardmodel}. First, one considers the action,
\bea
S_2[\phi,q,\pi] \; = \;  \frac{1}{2} \int dt \, \left( \dot{q}^2 + \alpha q^2 + \beta \phi^2  + 2 \pi (\dot\phi - q) \right) \, , 
\eea
whose equation for $\pi$ enforces the condition $q=\dot\phi$ that we substitute back to $S_2$ in order to recover $S_0$. Hence, we showed the equivalence between the two actions $S_0$ and $S_2$. 

As the action $S_2$ involves up to first order derivatives, we can perform the usual Hamiltonian analysis. We start by introducing  two pairs of conjugate momenta to parametrize the phase space,
\bea
\label{PS0}
\{q,p \} \, = \, 1 \, , \qquad \{ \phi, \pi \} \, = \, 1 \, .
\eea
We read directly from the action that $\phi$ and $\pi$ are conjugate whereas $p=\dot q$. Then, we compute the Hamiltonian and we immediately obtain the expression,
\bea
H \, = \, \frac{1}{2} \left( p^2 - \alpha q^2 - \beta \phi^2 \right) + \pi q \, .
\eea
In this point of view (i.e. from the Hamiltonian analysis of $S_2$), the Ostrogradsky instability shows up in the fact that $H$ is an affine function of $\pi$, and thus is unbounded  from below  and from above.  Such an analysis can be easily adapted to any higher derivative theory whose Lagrangian is at most quadratic in $\phi$ and its derivatives, and one can show that the corresponding Hamiltonian is still unbounded from below  and from above. 

Notice that, at this stage, the presence of the Ostrogradsky ghost in higher derivative theories does not mean that classical solutions are necessarily unstable. The theory \eqref{standardmodel} clearly admits a ghost but, depending on the coefficients $\alpha$ and $\beta$, its
solutions can be gentle oscillations about the vacuum $\phi=0$. As we are going to see later on in the section \ref{quantum}, the presence of the Ostrogradsky ghost becomes problematic when one adds self-interactions into the game and also in the quantum theory. Indeed, when the Hamiltonian is unbounded from below, the spectrum of the theory is unbounded from below as well and contains both positive and negative energy states. Hence, a self-interaction will produce a decay of the vacuum state into a collection of positive and negative energy states. 

\subsection{Generalization and Hamiltonian analysis}
Now, we are ready to study the more general higher derivative action 
\bea
\label{generalaction}
S[\phi] \; = \; \int dt \, L(\ddot\phi,\dot\phi,\phi) \, ,
\eea
where the Lagrangian $L$ is an arbitrary function of the three variables $\phi$, $\dot\phi$ and $\ddot\phi$, and to show how the Ostrogradsky shows up here.  First, we compute the corresponding  Euler-Lagrange equation,
\bea
\frac{d^2}{dt^2} \left( \frac{\partial L}{\partial \ddot \phi}\right) - \frac{d}{dt} \left( \frac{\partial L}{\partial \dot \phi} \right) + \frac{\partial L}{\partial \phi}\, = \, 0 \; ,
\label{eq:Lagrange_Equation_general}
\eea
which, after simple calculations,  reduces to the following equation of motion,
\bea
&&L_{\ddot\phi \ddot\phi} \, \ddddot \phi + L_{\ddot\phi \ddot\phi \ddot\phi} \, \dddot\phi^2 + L_{\ddot\phi \dot\phi \dot\phi} \, \ddot\phi^2
+L_{\ddot\phi \phi \phi} \, \dot\phi^2 +\nonumber \\
&& 2 (L_{\ddot\phi \ddot\phi \dot\phi} \dddot\phi \ddot\phi
+ L_{\ddot\phi \ddot\phi \phi} \,  \dddot\phi \dot\phi + L_{\ddot\phi \dot\phi \phi} \, \ddot\phi \dot\phi)  + (L_{\ddot\phi \phi} - L_{\dot\phi \dot\phi}) \, \ddot\phi - L_{\dot\phi \phi} \, \dot\phi + L_\phi \, = \, 0 \, .
\eea
For simplicity, we have introduced the standard notation $L_{xyz}$ for the derivatives of $L$ with respect to $x,y,z \in \{\ddot\phi,\dot\phi,\phi \}$.
This equation is  fourth order as in the quadratic case, and then, needs four initial conditions to be integrated. Hence, the theory propagates two degrees of freedom.

To show that one of these degrees of freedom is unstable, we perform a Hamiltonian analysis following the same method as in the previous
subsection. We   start by replacing \eqref{generalaction} by the equivalent first order action,
\bea
S_{\rm{eq}}[\phi,q,\pi]  \; = \;   \int dt \, \left( L(\dot q, q,\phi)   + \pi (\dot \phi - q)\right) \,.
\eea
The phase space is exactly the same as the one described in the previous subsection and can be parametrized by the two pairs of
conjugate variables \eqref{PS0}. Indeed, we read directly from the action that $\phi$ and $\pi$ are still conjugate whereas $p$ is given by,
\bea
\label{defofp}
p \; = \; \frac{\partial L}{\partial \dot q}(\dot q, q,\phi)  \, .
\eea
In order to compute the Hamiltonian, it is necessary to invert the previous equation so that we can express the velocity $\dot q$ in terms
of the phase space variables $p$, $q$ and $\phi$. In general, it cannot be done explicitly but, by virtue of the implicit function theorem, 
 it is possible to express locally, i.e. at the vicinity of any point ($\dot q_0$, $q_0$,$\phi_0$), the velocity $\dot q$ by solving \eqref{defofp}  provided that the ``non-degenerate Hessian'' condition,
\bea
\label{implicitcond}
\frac{\partial^2 L}{\partial \dot q^2}(\dot q_0,q_0,\phi_0) \neq 0 \, ,
\eea 
is satisfied, what we assume to be true in the whole phase space. 
In fact, such a condition is natural because it means that the initial Lagrangian \eqref{generalaction} does not reduce to an affine function in 
$\ddot \phi$ of the form $L(\ddot\phi,\dot\phi,\phi)=L_0(\dot\phi,\phi)+ \ddot \phi L_1(\dot\phi,\phi)$. If it was the case, the second derivative of
$\phi$ in $L$ could be easily eliminated with an integration by part using the relation,
\bea
\ddot \phi L_1(\dot\phi,\phi) = \frac{d}{dt}F(\dot\phi,\phi) - \dot \phi \frac{\partial F}{\partial \phi} \, , \qquad
F( \dot \phi,\phi) = \int d\dot\phi \, L_1(\dot\phi,\phi) \, ,
\eea
and thus the Lagrangian would reduce to an usual form with no higher derivatives. 

When \eqref{implicitcond} is fullfilled then, $\dot q$ can be expressed as a function of  $(p, q,\phi)$, and 
one can compute (implicitly) the Hamiltonian,
\bea
H \; = \;  \dot q(p, q,\phi)p - L(\dot q(p, q,\phi),q,\phi) + \pi q \, .
\eea
Hence, for any Lagrangian $L$ satisfying \eqref{implicitcond}, the Hamiltonian is an affine function of $\pi$ and thus it is unbounded from below and from above, and then an Ostrogradsky ghost is propagating inevitably. This proves the Ostrogradsky theorem for higher derivative theories. It can be easily generalized to Lagrangians with higher than second derivatives of $\phi$ by introducing further Lagrange multiplier terms replacing the higher derivatives of $\phi$ with new auxiliar variables as e.g. $q_2= \ddot \phi$ etc. as it was done recently in \cite{Motohashi:2018pxg}. 

At this stage, it seems impossible to avoid the Ostrogradsky  instability for higher derivative theories. However, there exist loopholes. Indeed,
in our analysis, we made many different assumptions that we can try to relax in order to evade the ghost. The first assumption is that the action  depends on one dynamical variable only. Does the conclusion of the Ostrogradsky theorem is affected when one considers theories with multiple variables? The second assumption is that we dealt so far with classical mechanics only and one may naturally wonder what happens
with field theories. We are going to address in details the first question concerning theories with multiple dynamical variables and, as it is now well known, we are going to show that the presence of higher derivatives does not necessary lead to an instability. We will shortly discuss the generalization to field theories, and more specifically to scalar-tensor theories, in the conclusion.

\section{Degeneracy or How to evade the Ostrogradsky ghost}
\label{sec:Degeneracy}
In this section, we show how to evade the Ostrogradsky ghost with degenerate higher-order Lagrangians of a coupled system, still
 in the context of classical mechanics.
 In a first subsection, we present the model we are going to study and we extend the Ostrogradsky theorem to Lagrangians of the form $L(\ddot\phi,\dot\phi,\phi;\dot x, x)$. Then, in a second 
subsection, we define the notion of degeneracy and show explicitly how it allows us to get rid of the Ostrogradsky ghost with a detailed
Hamiltonian analysis. Note, that the first two subsections are based on \cite{Motohashi:2016ftl}. Finally,  in the last subsection, we study the degeneracy conditions in order to find the general  properties that a Lagrangian 
$L(\ddot\phi,\dot\phi,\phi;\dot x, x)$ should have to be degenerate, and we show how the Euler-Lagrange equations reduce to a second
order system of two variables in that case. 

\subsection{Toy model: coupling to a regular extra variable}
In this section, to illustrate the importance of the notion of degeneracy in higher derivative theories, we consider the action of the form,
\bea
\label{Lphiq}
S[\phi,x] \; = \; \int dt \, L(\ddot\phi,\dot\phi,\phi ; \dot x , x) \, ,
\eea
which couples the two time-dependent variables $\phi$ and $x$. The Lagrangian contains at most second derivatives of $\phi$
whereas $x$ is a ``regular'' variable that appears at most with its first  derivative. Due to the presence of the second derivative $\ddot \phi$,
such theories exhibit, in general and without restrictions on $L$, an Ostrogradsky instability, which is compatible with  the fact that the equations of motion are higher order. Indeed, the Euler-Lagrange equations are given by,
\bea
&&\frac{\partial L}{\partial \phi} - \frac{d}{dt} \left( \frac{\partial L}{\partial \dot \phi} \right) + \frac{d^2}{dt^2} \left( \frac{\partial L}{\partial \ddot \phi}\right)  \, = \, 0 \; , \\
&&\frac{\partial L}{\partial x} - \frac{d}{dt} \left( \frac{\partial L}{\partial \dot x} \right) \, = \, 0 \, .
\eea
After some calculations,  one can easily show that they take respectively the general form,
\bea
&&L_{\ddot\phi \ddot\phi} \, \ddddot\phi  + L_{\ddot\phi \ddot\phi \ddot \phi} \,  \dddot\phi^2 + L_{\ddot \phi \dot x} \, \dddot x + 
{\cal A} \, \dddot \phi + {\cal B}   \; = \; 0 \, , \label{Eqforphi}\\
&&L_{\ddot\phi \dot x}\,  \dddot \phi + L_{\dot\phi \dot x} \, \ddot\phi + L_{\phi \dot x} \, \dot\phi + L_{\dot x \dot x} \, \ddot x + L_{\dot x x} \, \dot x - L_x \; = \; 0 \, , \label{Eqforx}
\eea
where $\cal A$ and $\cal B$ depend on the dynamical variables, their first and second derivatives only. Their expressions are easily computed but rather cumbersome
and not useful for our purposes. From these equations, it is not obvious to ``guess'' the necessary number of initial conditions to fully integrate them. However, on considering the time derivative of the equation for $x$ \eqref{Eqforx}, we obtain an equation very similar to the equation for $\phi$ \eqref{Eqforphi},
\bea
L_{\ddot \phi \dot x}\,  \ddddot\phi  + L_{\ddot\phi \ddot \phi \dot x} \, \dddot\phi^2  + L_{\dot x \dot x} \, \dddot x + {\cal C} \, \dddot\phi + {\cal D} \, \; = \; 0 \, ,
\eea
where  $\cal C$ and $\cal D$ depend on the variables ($x,\phi$), their first and second derivatives only. Then, we make use of this equation to 
solve $\dddot x$  and to substitute its expression in \eqref{Eqforphi} which finally becomes  equivalent to,
\bea
\left( L_{\ddot \phi \dot x}^2 - L_{\dot x \dot x} L_{\ddot\phi \ddot\phi}\right) \ddddot \phi + \left(L_{\ddot\phi \dot x} L_{\ddot\phi \ddot\phi \dot x} - L_{\dot x \dot x} L_{\ddot\phi \ddot\phi \ddot\phi} \right) \dddot\phi^2 + {\cal E} \, \dddot\phi + {\cal F} \; = \; 0 \, , \label{Eqforphi2}
\eea
where  $\cal E$ and $\cal F$ depend on the dynamical variables, their first and second derivatives only. As a consequence, the
equations of motion are equivalent to the system  \eqref{Eqforx} and \eqref{Eqforphi2} which is  at most second order in $x$ and can be up to fourth order in $\phi$. As a consequence, it is now clear that the equations of motion need, in general (without further conditions), six initial conditions in total to be integrated, corresponding to three physical degrees of freedom. One of these degrees of freedom is a ghost with an energy unbounded from below.

\medskip

To understand better the dynamics of  \eqref{Lphiq} and to prove the existence of a ghost in this theory, 
we proceed as in the previous section and we introduce the following equivalent first order action,
\bea
S_{\rm{eq}}[\phi,x,q,\pi] \; = \; \int dt \, 
\left( 
L(\dot q,q,\phi ; \dot x , x) + \pi (\dot \phi - q) 
\right) \; .
\eea
We can start a Hamiltonian analysis and we introduce the following three pairs of conjugate variables to parametrize the phase space,
\bea
\label{PSvariables}
\{q,p \}=1 \, , \qquad
\{ \phi, \pi\}=1 \, \qquad
\{x, p_x\}=1 \, .
\eea
It is clear from the action that $\phi$ and $\pi$ are conjugate whereas the two new momenta $p$ and $p_x$ are linked to the velocities by
the usual relations
\bea
\label{sytemPS}
p = \frac{\partial L}{\partial \dot q}(\dot q,q,\phi ; \dot x , x) \, , \qquad
p_x=\frac{\partial L}{\partial \dot x}(\dot q,q,\phi ; \dot x , x) \, .
\eea
To go further and to compute the Hamiltonian, we have to invert the system in order to express (if possible) the velocities $(\dot q, \dot x)$ in terms of the  phase space
variables, in particular the momenta $(p,p_x)$. In general, it is not possible to find an explicit and global solution to this problem. However, we
can invert locally (at the vicinity of any point in the phase space) the system. To show how this works more concretely, we first omit to mention 
the coordinate variables $(q,\phi,x)$ to lighten notations, and we assume that the system can be inverted for one point $(\dot q,\dot x)$ which
can thus be expressed in terms of the momenta $(p, p_x)$. Then, we linearize the system about this solution to see if it could extend to an open 
set at the vicinity of $(\dot q,\dot x)$. If we denote by $(\delta \dot q, \delta \dot x)$ the perturbations of the velocities, and by 
$(\delta p, \delta p_x)$ the perturbations of the momenta, the system \eqref{sytemPS} reduces to
\bea
\label{linearized}
\left(
\begin{array}{c} \delta p \\ \delta p_x \end{array}
\right) = 
K
\left(
\begin{array}{c} \delta \dot q \\ \delta \dot x \end{array}
\right) \, , \qquad 
K \equiv \left(
\begin{array}{cc}
L_{\dot q \dot q} &  L_{\dot q \dot x} \\
L_{\dot x \dot q}  & L_{\dot x \dot x} 
\end{array}
\right) \, ,
\eea
where the matrix $K$ is the so-called the kinetic matrix. We used the same notations as in the previous section to denote  partial derivatives of
$L$. Hence, it is possible to invert \eqref{linearized} if and only if $K$ is invertible. If this is the case at any point of the phase space, then 
\eqref{sytemPS} admits local solutions that can be glued together to form a global solution (which cannot be constructed explicitly in general)
in the whole phase space, 
\bea
\dot q (p,p_x,q,\phi,x) \, , \qquad
\dot x (p,p_x,q,\phi,x) \, .
\eea 
Furthermore, it becomes possible to define and compute the Hamiltonian given by,
\bea
\label{canonicalH}
H \; = \; p \dot q (p, p_x,q,\phi,x) + p_x \dot x (p, p_x,q,\phi,x) - L\left(\dot q(p, p_x,q,\phi,x),q,\phi ; \dot x(p,p_x,q,\phi,x) , x\right) + \pi q \, .
\eea
Hence, the system admits 3 degrees of freedom, and one sees immediately the linear instability which is, again, the signature of the
Ostrogradsky ghost. This result extends the Ostrogradsky theorem that we recalled in the previous section to the case where $\phi$ is
coupled to a regular variable $x$. Nonetheless, we saw here that the presence of the instability is intimately linked to the 
non-invertibility of the kinetic matrix $K$. Let us see now what happens when $K$ is degenerate, i.e. non-invertible. 

\subsection{Degenerate Lagrangians: Hamiltonian analysis}
In fact, the only hope to get rid of the linear instability in the theory \eqref{Lphiq} is to consider Lagrangians $L$ such that the kinetic
matrix is degenerate, i.e.
\bea
\label{degcond}
\text{det}(K) \; = \; L_{\ddot \phi  \ddot \phi} L_{\dot x \dot x} - L_{\ddot\phi \dot x}^2  \, = \, 0 \, ,
\eea
for any $\phi(t)$ and $x(t)$. Theories that satisfy this condition are called degenerate. Interestingly, we immediately remark that this condition
reduces the order of the equation of motion of $\phi$ \eqref{Eqforphi2} because the fourth derivative of $\phi$ comes with a term which
is exactly $\text{det}(K)$.

Furthermore, the degeneracy of $K$ changes drastically the conclusions of the Hamiltonian analysis. Indeed, in that case, the system of equations
\eqref{sytemPS} cannot be inverted, which means in practice that the momenta $p$ and $y$ are no more independent, and then they are
linked by a (primary) constraint. 

\subsubsection{Degeneracy and primary constraint}
To prove this statement precisely, we proceed in different steps and we treat different cases separately. 
In the first case, we assume that,
\bea
L_{\dot x \dot x} \neq 0 \Longleftrightarrow \frac{\partial y}{\partial \dot x} \neq 0 \, .
\label{case1}
\eea 
By virtue of the implicit function
theorem, one can express locally (at the vicinity of any point of the phase space), the velocity $\dot x$ in terms of $ p_x$, $\dot q$ and
the coordinates $(q,x,\phi)$. Hence, the momentum $p$ \eqref{sytemPS} can be expressed as 
$p = P(\dot q,p_x;q,x,\phi)$,
where $P$ is a function defined locally only. Furthermore, we are going to show that the degeneracy condition (that we have still not used so far) implies necessarily that $P$ does not depend on $\dot q$. If it was the case, i.e. if $\partial P/\partial \dot q \neq 0$, then (using again the implicit function theorem) one could express $\dot q$ in terms of the momenta $(p, p_x)$ (and the coordinates $(q,x,\phi)$) which contradicts the degeneracy condition. As a consequence, we are showing that the momenta $p$ and $p_x$ are linked by a primary constraint of the form,
\bea
\label{const1}
\chi \equiv p \, - \, P(p_x;q,x,\phi) \, \approx \, 0 \, ,
\eea
where $\approx$ is the usual notation for the weak equality, i.e. the equality up to the constraints in the phase space.

In the second case, we assume that
\bea
L_{\dot q \dot q} \neq 0 \Longleftrightarrow \frac{\partial p}{\partial \dot q} \neq 0 \, .
\label{case2}
\eea
Following exactly the same method as in the previous case, we show that there exists a function $Y$ in the phase space such that
\bea
\label{chi2}
\chi \equiv p_x \, - \, Y(p;q,x,\phi)  \, \approx \, 0 \, ,
\eea
which means, again, that there is a primary constraint in the theory. 

In the last case, we assume that the two previous conditions are not satisfied, i.e.,
\bea
L_{\dot q \dot q} = 0 =L_{\dot x \dot x} \, \Longleftrightarrow \,  \frac{\partial y}{\partial \dot x} = 0 =  \frac{\partial p}{\partial \dot q}  \, .
\label{case3}
\eea
Since the kinetic matrix $K$ is degenerate, it vanishes identically and then
\bea
\label{thirddegen}
L_{\dot x \dot q} = 0 \Longleftrightarrow  \frac{\partial y}{\partial \dot q} = 0 =  \frac{\partial p}{\partial \dot x}  \, .
\eea
As a consequence, neither $p$ nor $p_x$ depend on the velocity and the theory admits two primary constraints in that case,
\bea
\label{chipchiy}
\chi_{p_x} \, \equiv \, p_x \, - \, Y(q,x,\phi) \, \approx \, 0 , \qquad
\chi_p \, \equiv \, p \, - \, P(q,x,\phi) \, \approx \,  0 \, ,
\eea
where $Y$ and $P$ are functions of the coordinates $(q,x,\phi)$ only.

In summary, we conclude from this analysis that the degeneracy condition $\text{det}(K)=0$ is indeed equivalent to the existence of primary constraints
in the phase space. Depending on the dimension (one or two) of the kernel of $K$, the theory admits one or two primary constraints, which 
reduces the number of degrees of freedom and then could eliminate the Ostrogradsky ghost.

\subsubsection{Secondary constraint and the elimination of the Ostrogradsky ghost}

Let us pursue the Hamiltonian analysis in the first case of the previous subsection where the primary constraint is given by \eqref{const1}. Taking into account this constraint, the total
Hamiltonian  can be written as
\bea
\label{totalH}
H_T \; = \; H + \mu \chi \, = \, H_0 + \pi q + \mu \chi \, ,
\eea
where the canonical Hamiltonian $H=H_0 + \pi q$ has been defined in \eqref{canonicalH}, and $\mu$ is a Lagrange multiplier which enforces
the constraint $\chi \approx 0$. We are now ready to study the stability of the primary constraint under time evolution by computing,
\bea
\dot \chi & \equiv & \{ \chi, H_T \} = \{\chi, H\} = \{p  - 
P( p_x;q,x,\phi) , H_0(p,p_x;q,x,\phi) + \pi q\} \\
& = & - \pi + \Pi(p,p_x;q,x,\phi) \, ,
\eea
where $\Pi \equiv \{p  -  P , H_0\}$. Hence, we obtain a new (secondary) constraint,
\bea
\label{secondcons1}
\psi \equiv \pi - \Pi(p,p_x;q,x,\phi) \, \approx \, 0 \, ,
\eea
which fixes the momentum $\pi$ in terms of the remaining phase space variables. In other words, $\pi$ is no more an independent variable.
Generically, the two constraints $\chi$ and $\psi$ do not Poisson-commute, i.e.
\bea
\Delta \, \equiv \, \{ \chi , \psi \} \, \neq \, 0 \, ,
\eea
and then they form a set of second class constraints. Thus, requiring the stability under time evolution of $\psi \approx 0$ fixes the 
Lagrange multiplier $\mu$ and does not lead to a new (tertiary) constraint. This closes the Hamiltonian analysis. As we started with
6 degrees of freedom \eqref{PSvariables}, and we obtained a pair of second class constraints, we end up with 4 degrees of freedom in the phase space. 

The second case of the previous subsection is treated in a similar way. The total Hamiltonian is still given by \eqref{totalH} with the constraint $\chi$ \eqref{chi2}.
The requirement that the primary constraint is stable under time evolution leads (generically) to a secondary constraint given now by,
\bea
\psi \equiv \dot \chi = \frac{\partial Y}{\partial p} \pi - \frac{\partial Y}{\partial \phi} q + \{p_x-Y , H_0\} \, ,
\eea
which fixes also $\pi$ in terms  of the remaining phase space variables when ${\partial Y}/{\partial p}\neq 0$, otherwise $\pi$
is a free parameter. 
In general, the Hamiltonian analysis stops here with a pair of second class constraints and then four physical degrees of freedom 
in the phase space. 

Before studying the last case of the previous subsection, let us make some remarks on the first two cases which are very similar. It is now 
clear that the degeneracy condition leads to the elimination of (at least) one degree of freedom. Furthermore, 
we see that the secondary second class constraint $\psi \approx 0$ allows one in general
to express $\pi$ in terms of the phase space variables, and then if we substitute this expression back to the Hamiltonian, $H$ is no
more linear in $\pi$ and now there is no obvious reason to claim that $H$ is unbounded. 
However, at this stage, it is not obvious at all that we have eliminated the ghost.  
One needs a deeper study that can hardly be done in general to see whether the instability is still present or not. 

\medskip

Now, we return to the third case \eqref{thirddegen} which is different from the others. As $L_{\dot x \dot x}=L_{\ddot \phi \ddot \phi}=L_{\dot x \ddot\phi}=0$, it corresponds to a
Lagrangian at most linear in $\dot x$ and $\ddot\phi$, hence it is of the form,
\bea
L \; = \; \dot x Y(\dot\phi,x,\phi) + \ddot\phi P(\dot\phi,x,\phi) + Z(\dot\phi,x,\phi)  \, .
\eea 
The second derivatives of $\phi$ can be removed with an integration by part using the identity,
\bea
\ddot\phi P(\dot\phi,x,\phi) = \frac{d}{dt} Q(\dot \phi,x,\phi) - Q_x \dot x - Q_\phi \dot\phi \, , \qquad
Q(\dot\phi,x,\phi) \equiv \int d\dot\phi \, P(\dot\phi,x,\phi) \, .
\label{Lagcase3}
\eea
As a consequence, without performing any Hamiltonian analysis, we can conclude that the theory admits at most 2 degrees of freedom
with no Ostrogradsky ghost. In the Hamiltonian language, this translates into the fact that  $\chi_{p_x} \approx 0$ and $\chi_p \approx 0$
\eqref{chipchiy} form (generically) a set of second class constraints and then kill one degree of freedom. In the particular case where
these two constraints commute, the theory admits either a gauge symmetry or more constraints which eliminate even more degrees
of freedom, leading to a theory with less than one degree of freedom. 

\subsection{Degenerate Lagrangians:  explicit forms and equations of motion}
As one could have expected,  the higher-order Euler-Lagrange equations of a degenerate Lagrangian can be reduced to
a system of two second order equations of motion, what we are going to show now. For that purpose, it is very convenient to first 
compute the solutions of the 
constraint \eqref{degcond} for the Lagrangian $L$. 

We have already solved the constraint in the case \eqref{case3} where it was shown that the Lagrangian
reduces to the form \eqref{Lagcase3} and then does not involve any  higher order derivatives. Hence, the two equations of motion 
for $\phi$ and $x$ are obviously second order.

The constraint is much more involved to solve in the first two cases \eqref{case1} and \eqref{case2}. We will mainly focus on the first case as the second one can be treated in a very similar way. Hence, the degeneracy constraint becomes equivalent to \eqref{const1},
\bea
\label{EqLcase1}
\frac{\partial L}{\partial \dot q} - P(\frac{\partial L}{\partial \dot x};q,x,\phi) \; = \; 0 \, ,
\eea
which is a non-linear partial differential equation for the Lagrangian $L$, viewed as a function of $(\dot q, \dot x)$. The variables $(q,x,\phi)$
play the role of parameters in this equation, and then we will omit to mention them in what follows to lighten  notations. 

\subsubsection{General implicit solution and second order equations}

To solve such an equation, one can use the method of characteristics, and first transform the non-linear partial differential equation into a quasi-linear partial differential equation as follows,
\bea
\label{EqLdercase1}
\frac{\partial u}{\partial \dot q} - P'(u) \frac{\partial u}{\partial \dot x} \; = \; 0 \; , \qquad
u \equiv \frac{\partial L}{\partial \dot x} \, ,
\eea  
where $P'$ is the derivative of $P$ with respect to $ p_x \equiv {\partial L}/{\partial \dot x}$.
More precisely, as one obtains the previous equation \eqref{EqLdercase1} by deriving  \eqref{EqLcase1} with respect to $\dot x$, then 
\eqref{EqLdercase1} is in fact only a necessary condition for \eqref{EqLcase1} to be satisfied. General solutions of \eqref{EqLdercase1} 
are well-known and are given by the implicit relation,
\bea
\label{generalsolofPDE}
\dot x \; = \; - \dot q \, P'(u) + Q(u) \, ,
\eea
where $Q$ is an arbitrary function of $u$, and also $(q,x,\phi)$ which have not been mentioned for simplicity. An explicit solution for $u$
does not exist in general when $P$ and $Q$ are arbitrary functions. However, there exist some situations where we can solve 
$u$ and find the Lagrangian explicitly.

One of the simplest situation corresponds to the case where $P'(u)=P_0$ is a constant, i.e. it does not depend on $u$ but it could depend on the variables $q$, $x$ and $\phi$. In that case, the constraint becomes a linear partial differential equation. Hence, $u$ is easily found and reads,
\bea
u \; = \; Q^{-1}(\dot x + P_0 \dot q) \, ,
\eea
where $Q^{-1}$ is the inverse of $Q$ for the composition rule, i.e. $Q(Q^{-1}(u))=u$. As a consequence,  the Lagrangian $L$
which is related to $u$ by \eqref{EqLdercase1} is given by,
\bea
\label{solL0}
L \; = \; F(\dot x + P_0 \dot q) + C(\dot q) \, ,
\eea
where $F'=Q^{-1}$ and $C$ is a function that does not depend on $\dot x$ and shows up because of the integration of $u$.
As we recalled previously, solutions of the degeneracy conditions \eqref{EqLcase1} are necessary of the form \eqref{solL0}, but
this is not sufficient. We have to impose that \eqref{solL0} is indeed a solution of \eqref{EqLcase1}, and we easily show that this is the
case if and only of $C$ is a constant. Finally, if we  express again all the variables in $L$ and replace $q$ by $\dot \phi$, we conclude that 
the general solution of the degeneracy condition for the Lagrangian takes the form,
\bea
\label{solL1}
L(\ddot \phi,\dot\phi,\phi;\dot x,x) \; = \; F ( \dot x + P_0(\dot\phi,\phi,x) \ddot\phi ; \dot \phi,\phi, x ) \, , 
\eea 
where $F$ is an arbitrary function. Notice that one could have found this solution directly from solving the equation \eqref{EqLcase1} when
\bea
P(\frac{\partial L}{\partial \dot x};q,x,\phi) \; = \; P_0(q,x,\phi) \, \frac{\partial L}{\partial \dot x} \, ,
\eea
without going through the general solution of the non-linear equation. 

One important consequence of \eqref{solL1}
is that the second derivative of $\phi$ can be totally absorbed thank to a  change of variable. Instead of $x$,
we use the variable $X$ defined by,
\bea
\label{defX}
X \; \equiv \; x + R(\dot\phi,\phi,x) \, , 
\eea
where $R$ satisfies the partial differential linear equation,
\bea
\frac{\partial R}{\partial \dot\phi} - P_0 \frac{\partial R}{\partial x} = P_0 \, ,
\eea
which can be shown to have a solution. The relation \eqref{defX} enables us to express $x$ in terms of $X$, and the partial differential
equation allows to get rid of $\ddot\phi$ in the Lagrangian using the identity,
\bea
\dot X \; = \; (1+ \frac{\partial R}{\partial x}) ( \dot x + P_0 \ddot \phi ) + \frac{\partial R}{\partial \phi} \, \dot\phi \, .
\eea
As a consequence, the Lagrangian \eqref{solL1} can reformulated in terms of $X$, $\phi$ and its first derivatives only. Therefore, the two equations of motion are necessarily second order and the system admits, as expected, only two degrees of freedom.

As we have already emphasized, in general, an explicit solution of the degeneracy condition for the Lagrangian is not known. However,
it is possible to show that the degeneracy condition is enough to  prove that the Euler-Lagrange equations for $\phi$ and $x$ can be
reformulated as a system of two second order differential equations. This is consistent with the Hamiltonian analysis which shows that 
there is no extra degree of freedom in these theories. Nonetheless, it is worth stressing that the Euler-Lagrange equations derived from the Lagrangian in general do not give directly the ``minimal'' system of equations because they can involve up to fourth-order derivatives of $\phi$, as we saw. Demanding the Euler-Lagrange equations to be second order is clearly not a necessary requirement in order to avoid the Ostrogradsky ghost.

Let us quickly mention that the study of degenerate theories can be generalized for Lagrangians of
the form $L(\ddot\phi_a,\dot\phi_a,\phi_a;\dot x_i,x_i)$ where there is an arbitrary number of $\phi$-like variables and 
regular ($x$-like) variables in  \cite{Motohashi:2016ftl} (see also \cite{Motohashi:2014opa}), and  for (multi-) scalar fields theories in a flat space-time \cite{Crisostomi:2017aim}. 
The notion of degeneracy has also been generalized to theories with third derivatives of the dynamical variables in \cite{Motohashi:2017eya}.

\subsubsection{Examples}

To conclude this section, we propose to give some examples to make the results we have just presented more concrete. Let us start
with a simple generalization of \eqref{standardmodel}  whose action reads,
\bea
\label{quaddegaction}
S[\phi,x] \; =  \; \frac{1}{2} \int dt \, \left( \lambda  \ddot{\phi}^2 + \alpha \dot \phi^2 + \beta \phi^2  + \kappa  \dot x^2 + \gamma  x^2 
+2 \mu \dot x \ddot \phi \right) \, ,
\eea
which involves a higher derivative coupling of $\phi$ to an extra regular variable $x$. 
All the coefficients entering in the action are supposed to be constant. Such a model has been considered in \cite{Langlois:2015cwa}.
If the theory is degenerate, the constraint is necessarily
linear because the  theory is free, i.e. quadratic in the dynamical variables. Hence, according to what we have shown \eqref{solL1}, the theory
is degenerate if and only the ``kinetic'' part of the Lagrangian factorizes according to,
\bea
 \lambda  \ddot{\phi}^2 + \kappa \dot x^2 +  2 \mu \dot x \ddot \phi = \kappa (\dot x + P \ddot\phi)^2 \, ,
\eea
which  is possible if and only if,
\bea
\lambda \kappa - \mu^2 \; = \; 0 \; , \qquad P = \frac{\mu}{\kappa} =\frac{\lambda}{\mu}\, .
\eea
Following \eqref{defX}, we change variable and instead of $x$, we use the variable $X \equiv x + P \dot\phi$, so that the action 
\eqref{quaddegaction} simplifies and becomes,
\bea
S[\phi,X] \; = \; \frac{1}{2} \int dt \, \left( \kappa \dot X^2 + \alpha \dot\phi^2 + \beta \phi^2 + \gamma (X-P \dot\phi)^2 \right)\, .
\eea
It is now clear that the theory propagates two degrees of freedom only, it has no Ostrogradsky ghost and the equations of motion are
second order. Notice that the theory could suffer from other types of instability if the kinetic matrix is not positive definite for instance.

\medskip

Another interesting example which does not fall in the class studied before is given by a theory whose kinetic part of the Lagrangian is factorized as follows,
\bea
L \, = \, A(\ddot\phi;\dot\phi,\phi,x) B(\dot x;\dot\phi,\phi,x) + C(\dot\phi,\phi,x) \, .
\eea
The kinetic matrix is easily computed,
\bea
 K \; = \; \left(
 \begin{array}{cc}
 A_{\ddot \phi \ddot\phi} B & A_{\ddot \phi} B_{\dot x} \\
 A_{\ddot \phi} B_{\dot x} & A B_{\dot x \dot x}
 \end{array}
 \right)
\eea
and the degeneracy condition reduces to two ordinary differential equations,
\bea
A A_{\ddot \phi \ddot\phi}  = \lambda (A_{ \ddot\phi} )^2 \, , \qquad
B B_{\dot x \dot x} = \frac{1}{\lambda} (B_{\dot x})^2 \, ,
\eea
where $\lambda$ is a function of $(\dot\phi,\phi,x) $ only. One can easily integrate these equations and, assuming $\lambda \neq 1$, one
obtains the solutions,
\bea
A \; = \; (\mu_A \ddot\phi + \nu_A)^{\alpha} \, , \qquad
B \; = \; (\mu_B \dot x + \nu_B)^{1-1/\alpha} \, , \qquad 
\alpha \; = \;  \frac{1}{1-\lambda} \, ,
\eea
where $\mu_X$ and $\nu_X$ for $X \in \{A,B\}$ are functions of $(\dot\phi,\phi,x)$. 
When $\alpha=-1$, one recovers the well-known
degenerate Lagrangian,
\bea
L \; = \; \frac{\left(\mu_B(\dot\phi,\phi,x) \dot x + \nu_B(\dot\phi,\phi,x)\right)^2}{\mu_A(\dot\phi,\phi,x)  \ddot\phi + \nu_A(\dot\phi,\phi,x)} + C(\dot\phi,\phi,x)\, .
\eea
In this example, we see that, contrary to the previous example, 
the degeneracy condition is not linked to the possibility of absorbing higher derivatives in a change of variable.
More examples can be found in \cite{Motohashi:2016ftl}.

\section{Quantization of the Ostrogradsky ghost}
\label{quantum}
This section is devoted to illustrate how the Ostrogradsky instability shows up at the quantum level.
In that aim, we consider free theories, whose Lagrangian is quadratic in the dynamical variables, 
that we can quantize completely. In particular, we construct the Hilbert space of quantum states and we see
that the presence of an Ostrogradsky ghost in the classical theory leads necessarily to a problem in the quantum
theory which can be an instability, the presence of negative norm states or  non-normalizable states, as it was shown in 
\cite{Smilga:2004cy,Smilga:2017arl}. As quantum instabilities can be  related to classical instabilities, 
we will also  study the classical stability properties of free theories which propagate an Ostrogradsky ghost.
For instance, the quantization of a free theory with complex frequencies (and exponentially growing classical solutions) 
leads to non-unitarity problems. 
Once we identify the free theories which are classically stable, we will study their quantization. 

In the first two subsections, which are based on \cite{Smilga:2017arl,Smilga:2004cy}, we review the classical and quantum properties of the Pais-Uhlenbeck model \cite{Pais:1950za} which provides
the most general quadratic Lagrangian  $L(\ddot \phi, \dot \phi, \phi)$ with oscillating (and not exponentially increasing or decreasing) 
classical solutions. In the last subsection, we extend the analysis by coupling the  Pais-Uhlenbeck model  to a regular variable and we show that, without
degeneracy, the theory has  classical and quantum instabilities.

\subsection{The Pais-Uhlenbeck model with different frequencies}
As we have said, our aim is to study the family of theories whose dynamics is governed by the action \eqref{quaddegaction}. Without loss of generality, 
we can fix $\lambda=1$ as we are not considering degenerate theories here,
\bea
\label{fulltheory}
S[\phi,x] \; =  \; \frac{1}{2} \int dt \, \left( \ddot{\phi}^2 + \alpha \dot \phi^2 + \beta \phi^2  + \kappa  \dot x^2 + \gamma  x^2 
+2 \mu \dot x \ddot \phi \right) \, .
\eea

From the analysis in the previous section, we know that, as the theory is non-degenerate, it propagates three degrees of freedom. 
The equations of motion are easily computed and read,
\bea
\ddddot\phi - \alpha \ddot\phi + \beta \phi + \mu \dddot x = 0 \, , \qquad
\kappa \ddot x - \gamma x + \mu \dddot\phi = 0 \, .
\eea
It is clear that there is only one equilibrium point at $x=0$ and $\phi=0$. Now, we ask the question whether this
point is stable or not. In fact, we require an even stronger condition and ask that the point is stable even in the absence of coupling between
$\phi$ and $x$ (when $\mu=0$). Hence,  the variable $x$ must be an harmonic oscillator which implies the conditions,
\bea
\label{omega0}
\kappa > 0 \, , \qquad \gamma \equiv  -\kappa \omega_0^2 < 0 \, ,
\eea
where $\omega_0$ is the frequency.

\subsubsection{The classical theory: decoupling of the two modes}
Let us now consider  the theory for $\phi$ only where there is no regular variables, i.e. $x=0$ identically . We look for oscillating solutions of the form $ \phi \propto \exp(i \omega t)$ and we
immediately obtain the constraint that the roots of the equation,
\bea
\omega^4 + \alpha \omega^2 + \beta \; = \; 0 \, ,
\eea
must be real and positive. If we denote $\pm \omega_1$ and $\pm \omega_2$ its roots, then
$\alpha=-(\omega_1^2 + \omega_2^2)$ and $\beta=\omega_1^2\omega_2^2$. Hence, we impose the conditions
\bea
\label{alphabeta}
\alpha= -(\omega_1^2 + \omega_2^2) <0 \, , \qquad
\beta=\omega_1^2\omega_2^2 >0 \, ,
\eea
which are necessary and sufficient for the theory to have a stable vacuum.
Therefore, we recover, as expected, the famous Pais-Uhlenbeck Lagrangian \cite{Pais:1950za},
\bea
\label{PUaction}
L_{\rm{pu}_0} \; \equiv \; \frac{1}{2} \left(\ddot\phi^2 - (\omega_1^2+\omega_2^2) \, \dot\phi^2 + \omega_1^2\omega_2^2 \, \phi^2 \right)  \, ,
\eea
 where $\omega_1$ and $\omega_2$
are the frequencies of the two modes propagating in the theory.  

Notice that, even when the theory is classically stable, 
the ghost still propagates. Thus, in general, there is no direct link between the presence of an Ostrogradsky ghost in a free theory and the stability
properties of classical solutions. The reason why we are interested here in theories with stable classical solutions is that these theories are
the only one that could provide a quantum theory with no problem of unitarity. In other words, if the theory has unstable classical solutions,
the quantum theory will not be unitary.

At this stage, one may ask the question which mode in the Pais-Uhlenbeck model is the Ostrogradsky ghost while the Lagrangian 
is totally symmetric in the exchange of the frequencies $\omega_1$ and $ \omega_2$. To answer this question, it is useful
to reformulate the action \eqref{PUaction} in such a way that the two modes decouple. Hence, we start by replacing the Lagrangian 
by its ``first order'' equivalent form,
\bea
\label{Lpu2}
L_{\rm{pu}_1} \; = \; -  \dot\phi \dot\psi  - \frac{1}{2} \left(
\psi^2  + (\omega_1^2+\omega_2^2) \, \dot\phi^2 - \omega_1^2\omega_2^2 \, \phi^2 
\right) \, ,
\eea
from which we quickly deduce the equations of motion that can be written in a matrix form as follows,
\bea
\left(
\begin{array}{c} \ddot\psi \\ \ddot \phi \end{array}
\right) + 
\left(
\begin{array}{cc} \omega_1^2 + \omega_2^2 & \omega_1^2 \omega_2^2 \\ -1 & 0 \end{array}
\right)
\left(
\begin{array}{c} \psi \\ \phi \end{array}
\right) = 0 \, .
\label{matrixPU}
\eea
To disentangle the system of equations for $\phi$ and $\psi$, we have to diagonalize the $2 \times 2$ matrix above \eqref{matrixPU}. One sees immediately that
$\omega_1^2$ and $\omega_2^2$ are the eigenvalues while the eigenvectors are,
\bea
\label{changex1x2}
x_1 = \omega_2^2 \phi + \psi \, , \qquad x_2 = \omega_1^2 \phi + \psi \, ,
\eea
which satisfy the standard harmonic oscillator equation $\ddot x_a + \omega_a^2 \, x_a =0$ for $a \in \{1,2\}$. As a consequence, 
if we reformulate \eqref{Lpu2} in terms of $x_1$ and $x_2$, we obtain a decoupled Lagrangian $L_{\rm{pu}_2} $ given by,
\bea
\label{decoupledLag}
L_{\rm{pu}_2} \; = \; \frac{1}{2(\omega_1^2 - \omega_2^2)} \left[ (\dot x_1^2 - \omega_1^2 \, x_1^2) - (\dot x_2^2 - \omega_2^2 \, x_2^2)  \right] \, .
\eea
We first remark that the decoupling is possible only if the frequencies are different, $\omega_1 \neq \omega_2$; the case where 
the frequencies are equal will be discussed later. Interestingly, we see that the Pais-Uhlenbeck 
Lagrangian can be written as the difference of two harmonic oscillator Lagrangians and the ghost corresponds to the mode
with lowest  frequency, say $\omega_2$. Thus, there is a kind of ``spontaneous'' symmetry breaking because, even though the action
is symmetric under the exchange of the frequencies, the lowest one is a ghost while the highest one is a regular harmonic oscillator. 

\subsubsection{Self-interaction and classical instability}

\begin{figure}
\includegraphics[width=6cm,height=6cm]{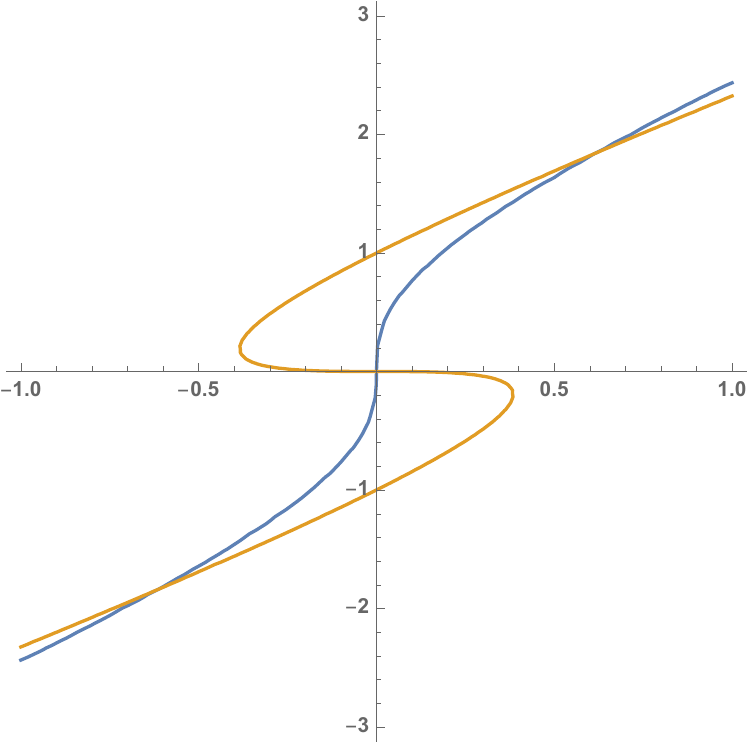}\
\caption{The curves $F_1(x_1,x_2)=0$ (blue curve) and $F_2(x_1,x_2)=0$ (orange curve) where $x_1$ is the horizontal axis and $x_2$ the vertical one.  
We fixed the parameters to $\lambda=1$, $\omega_1^2=3$ and $\omega_2^2=1$.
Above the blue curve, $F_1>0$ and on le left of the orange curve, $F_2>0$. The intersections of the two curves give the three vaccums.
We see that, at the vicinity of these points, there are always regions where, at least, one of the strengths is positive which implies that the vaccums are all unstable.}
\label{plots}
\end{figure}

As we are going to see, the presence of such a ghost makes the quantum theory unstable. Indeed, the spectrum
is clearly unbounded from below with positive energy and negative energy states, and the introduction of a self-interaction in 
the theory will destabilize the vacuum state which will decay into positive and negative energy states. At the classical level, one can already see 
that a self-interaction makes the vacuum unstable as shown in \cite{Smilga:2017arl}. To see this is the case, let us add to the free Lagrangian \eqref{PUaction}, a self
interaction term  $ L_{\rm{self}} \propto \phi^4$. After the change of variable \eqref{changex1x2}, the self-interaction can be reformulated as
follows,
\bea
L_{\rm{self}} \; = \;  \frac{\lambda}{8(\omega_2^2 - \omega_1^2)} (x_1-x_2)^4 \, ,
\eea
where $\lambda$ is a coupling constant. The normalization factor has been chosen in order to simplify the equations of motion which become,
\bea
&&\ddot x_1 - F_1 (x_1,x_2) \equiv \ddot x_1 + \omega_1^2 x_1 + \lambda (x_1-x_2)^3 = 0 \, , \\
&&\ddot x_2 - F_2 (x_1,x_2) \equiv \ddot x_2 + \omega_2^2 x_2 - \lambda (x_2-x_1)^3 = 0 \, .
\eea
There are now three classical vacua, the first one is the same as the previous one $x_1=x_2=0$ (without self-interaction) and the two others are given by,
\bea
\label{neweq}
x_{10}^2= \frac{\omega_1^2 \omega_2^6}{\lambda(\omega_1^2-\omega_2^2)^3} \, , \qquad
x_{20}= \frac{\omega_1^2}{\omega_2^2} x_1 \, ,
\eea
which exist only if $\lambda\,(\omega_1-\omega_2)>0$, what we assume to be the case (in fact, we assume $\omega_1>\omega_2$ and then $\lambda>0$). The first point $x_1=x_2=0$ is now unstable, as it is show in  Fig. (\ref{plots}).

Concerning the  new points, we expand the equations of motion about these points \eqref{neweq} and we obtain the linear system,
\bea
\left(
\begin{array}{c}
\ddot \delta_1 \\ \ddot \delta_2
\end{array}
\right) +
\left(
\begin{array}{cc}
\omega_1^2 + \omega_3^2 & - \omega_3^2 \\
\omega_3^2 & \omega_2^2 - \omega_3^2
\end{array}
\right) 
\left(
\begin{array}{c}
\delta_1 \\ \delta_2
\end{array}
\right)=0 \, , \qquad 
\omega_3^2 \equiv  \frac{3\, \omega_1^2 \omega_2^2}{\omega_1^2 - \omega_2^2 }
\, .
\eea
As the  $2\times 2$ matrix in  this equation  is such that its determinant is negative,
\bea
\left\vert 
\begin{array}{cc}
\omega_1^2 + \omega_3^2 & - \omega_3^2 \\
\omega_3^2 & \omega_2^2 - \omega_3^2
\end{array}
\right\vert = -2 \omega_1^2 \, \omega_2^2 \, ,
\eea
it admits one positive and one negative eigenvalue, and then the system is unstable.
This is a consequence of the fact that the energy of the free system is unbounded from below. If we had started with
two well-defined harmonic oscillator, we would have found two stable points.

\subsubsection{Quantization: unbounded spectrum and negative norme states}
\label{quantumPU}
Let us now say a few words on the canonical quantization.
As we assumed that $\omega_1 > \omega_2$, we  can introduce the mass $m \equiv (\omega_1^2-\omega_2^2)^{-1}$, and the Hamiltonian of the theory \eqref{decoupledLag} reads,
\bea
H \; = \; \left( 
\frac{p_1^2}{2 m} + \frac{m \omega_1^2 x_1^2}{2} \right)-  \left(\frac{p_2^2}{2m} + \frac{m \omega_2^2 x_2^2}{2}
\right) \, ,
\eea
where $\{x_1,p_1\}=1=\{x_2,p_2 \}$ in the phase space. 

The quantization is straightforward. Indeed, the two pairs of conjugate variables become non-commutative operators,
\bea
\label{quantumalgebra}
[\hat{x}_1,\hat{p}_1] = i \hbar = [\hat{x}_2,\hat{p}_2] \, ,  
\eea
where $\hbar$ is the Planck constant, and the quantum states $\vert n_1,n_2 \rangle$ are labelled by two non-negative 
integers $(n_1,n_2)$. These states from a basis of a Hilbert space and are eigenvectors of the Hamiltonian operator
with eigenvalues,
\bea
E(n_1,n_2) \; = \; \frac{\hbar \omega_1}{2} \left( n_1 + \frac{1}{2} \right) \, 
- \, \frac{\hbar \omega_2}{2} \left( n_2 + \frac{1}{2} \right) \, .
\eea
The spectrum is obviously unbounded and the introduction of self-interactions will destabilize the theory. 
Notice that one can render the spectrum positive at the price to have negative norm states. In any case, the quantum theory
is sick as it was clearly shown in \cite{Smilga:2017arl}. 

\subsection{The Pais-Uhlenbeck model with equal frequencies}
At this stage, the only hope to ``save'' the Pais-Uhlenbeck model from quantum and classical instabilities is to consider theories 
with equal frequencies, i.e. $\omega_1=\omega_2 = \omega$. 

\subsubsection{The classical theory: linear instability}

This case
is very different from the previous one mainly because the two equations of motion of \eqref{Lpu2}, which now become,
\bea
\left(
\begin{array}{c} \ddot\psi \\ \ddot \phi \end{array}
\right) + 
\left(
\begin{array}{cc} 2 \omega^2 & \omega^4  \\ -1 & 0 \end{array}
\right)
\left(
\begin{array}{c} \psi \\ \phi \end{array}
\right) = 0 \, ,
\eea
cannot be decoupled. Indeed, the $2\times 2$ matrix that appears in these equations of motion is no more diagonalizable, which can be
easily seen if one makes the simple change of variable $\phi \rightarrow \varphi \equiv \omega^2 \phi + \psi$, 
so that the equations of motion become,
\bea
\left(
\begin{array}{c} \ddot\psi \\ \ddot \varphi \end{array}
\right) + \omega^2
\left(
\begin{array}{cc} 1 & 1  \\ 0 & 1 \end{array}
\right)
\left(
\begin{array}{c} \psi \\ \varphi \end{array}
\right) = 0 \, .
\eea
As the matrix is triangular, it is now clear that it is not possible to decouple the dynamics of the two degrees of freedom. 
Nonetheless, the resolution of the equations of motion is trivial and gives,
\bea
\varphi(t) = A \cos(\omega t + \Phi) \, , \qquad
\psi(t) = B  \cos(\omega t + \Psi) -  \frac{ A t}{2\omega} \sin(\omega t + \Phi) \, ,
\eea
where $A,B,\Phi$ and $\Psi$ are the four integration constants. The fact that $\psi$ has a term linear in $t$  makes
the vacuum unstable. Even if the instability is ``softer" than the  usual exponential instability,
it could have important consequences in the quantum theory. 

 As the theory is not decoupled, the quantization will not be as simple as  the case
of unequal frequencies. To prepare the quantization, it is interesting to consider the following theory,
\bea
S[x_1,x_2] = \frac{1}{2}\int dt \, \left( \dot x_1^2 + \dot x_2^2 + \omega^2(x_1^2 + x_2^2) - 2\omega \, (\dot x_1 x_2 - x_1 \dot x_2) \right) \, ,
\eea
whose equations of motion are,
\bea
\ddot x_1 - \omega^2 x_1 - 2\omega \dot x_2 \, = \, 0 \, , \qquad
\ddot x_2 - \omega^2 x_2 + 2\omega \dot x_1 \, = \, 0 \, .
\eea
This system of equations can be easily shown to be  equivalent to the system,
\bea
\ddddot x_2 + 2 \omega^2 \, \ddot x_2 + \omega^4 \, x_2 \, = \, 0 \, , \qquad
x_1 = -\frac{1}{2 \omega^3} (3 \omega^2 \, \dot x_2 + \dddot x_2) \, ,
\eea
which proves the equivalent with the Pais-Uhlenbeck theory with $\omega_1=\omega_2=\omega$ whose equations of motion
are exactly the same. What makes this reformulation
of the Pais-Uhlenbeck theory particularly interesting is the shape of its Hamiltonian. Indeed, if we introduce the two pairs of 
conjugate variables in the phase space,
\bea
\label{classicalPoi}
\{ x_1,p_1 \} \, = \, 1 \, = \, \{ x_2,p_2 \} \, ,
\eea
then we easily compute the Hamiltonian and we find that it takes the form,
\bea
\label{HamilL}
H \; = \; \frac{1}{2}(p_1^2+p_2^2) + \omega L \, , \qquad L \equiv p_1 x_2 - p_2 x_1 \, ,
\eea
where $L$ is the two-dimensional angular momentum. The form of this Hamiltonian will make the quantization much easier. This result was originally found in \cite{Smilga:2017arl} in a different way (from a canonical transformation).

\subsubsection{Quantization: non-normalizable quantum states}
To quantize the theory, we follow \cite{Smilga:2004cy} and we start by replacing the classical Poisson algebra \eqref{classicalPoi} by its quantum
counterpart  \eqref{quantumalgebra}. The unique unitary irreducible representation of this quantum algebra is the space
$L^2(\mathbb R^2,\mathbb C)$ of square integrable (complex valued) functions of two real variables. 

Now, we look for the eigenstates and eigenvalues of the quantum Hamiltonian 
\bea
\hat{H} \; = \; -\frac{\hbar^2}{2} \Delta + \omega \hat{L} \, ,
\eea 
where $\Delta$ is the 2-dimensional Laplacian and $\hat{L}$ the quantum angular momentum. Interestingly $\Delta$ and $\hat{L}$
commute, hence the eigenstates of $\hat{H}$ are eigenstates of both $\Delta$ and $\hat{L}$.  The eigenstates of the angular momentum operator $Y_\ell(r,\theta)$ are labelled by an integer $\ell \in \mathbb Z$ and they are
easily written in cylindrical coordinates $(r,\theta)$ as follows, 
\bea
Y_\ell(r,\theta) = R(r) \exp (i \ell \theta) \, , \qquad \hat{L} Y_\ell  = -i\hbar \frac{\partial Y_\ell}{\partial \theta} = \hbar \ell \, Y_\ell\, ,
\eea
where $R(r)$ is still an arbitrary function of $r$ at this stage.  Then, the eigenvalue equation of $\hat{H}$ translates into a differential equation for $R(r)$ given by,
\bea
-\frac{\hbar^2}{2} \left( R'' + \frac{1}{r} R' - \frac{\ell^2}{r^2} R\right) + \hbar \omega \ell R =  E \, R \, , 
\eea
where $E$ is the energy of the system, and $R'$ (resp. $R''$) denotes the first (resp. second) derivative with respect to $r$. If we introduce the notations,
\bea
x \; \equiv \; k \, r \, , \qquad
k^2 \equiv \frac{2}{\hbar^2}(E-\hbar \omega \ell) \, ,
\eea
then the previous equation becomes clearly a Bessel equation,
\bea
x^2 \frac{d^2 R}{dx^2} + x \frac{dR}{dx} + (x^2-\ell^2) R \, = \, 0 \, ,
\eea
whose solution is $R(x)=J_\ell(x)$, or equivalently $R(r)=J_\ell(k r)$ where $k$ is a priori a complex number and $J_\ell$ the Bessel function
of the first kind. 
Finally, the eigenstates
$\Psi_{\ell,E}$ of the Hamiltonian operator are labelled by a pair $(\ell,E)$ where $\ell$ is an integer and $E$ a real number,
\begin{align}
\label{quantumstates}
\Psi_{\ell,E}(r,\theta) \; =\; J_{\ell} \left( \frac{r}{\hbar} \sqrt{2(E-\hbar \omega \ell)} \right) \, \exp(i \ell \theta) \, .
\end{align}
As a consequence, the spectrum is continuous with no ground state and the quantum eigenstates of the energy are non-normalizable (with
the usual measure on $\mathbb R^2$), i.e. 
\bea
\int r\,dr \, d\theta \,  \vert \Psi_{\ell,E}(r,\theta)  \vert^2 \, = \, \infty \, .
\eea
All this makes the
quantum theory ill-defined even though the Hilbert space can be formally constructed by asking that $\Psi_{\ell,E}$ are orthonormal states
for instance. But, in that case, we are loosing the interpretation of $(r,\theta)$ as a radial and angular variables, and the operators $x_a$ and
$p_a$ are no more unitary.

\subsection{Coupling to a regular variable}
\begin{figure}
\includegraphics[width=6cm,height=6cm]{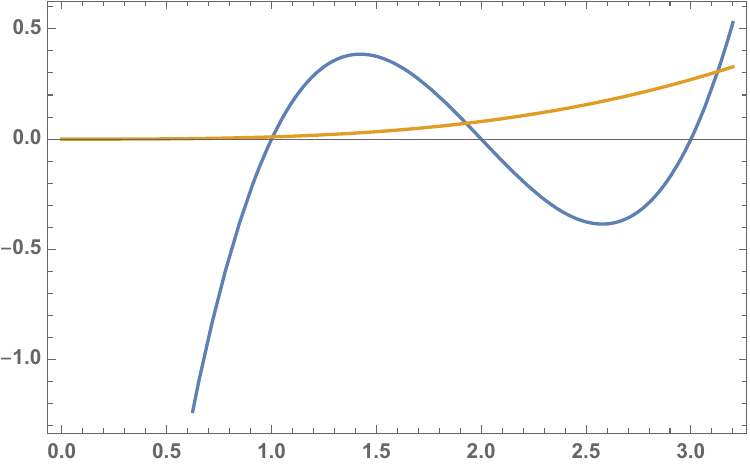}\
\includegraphics[width=6cm,height=6cm]{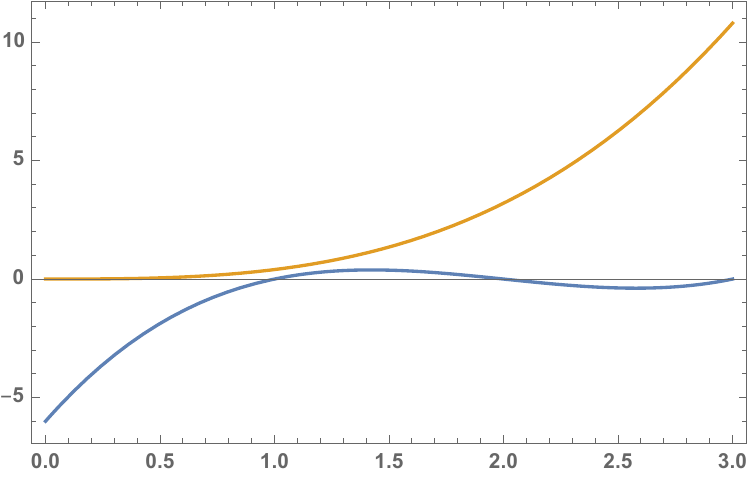}\
\caption{The two curves $y=(\omega^2-\omega_1^2) (\omega^2-\omega_2^2)(\omega^2-\omega_3^2)$ and $y=4\mu^2/\kappa\omega^6$ (for particular values of the parameters) as functions of $\omega^2$. On the l.h.s., $\mu^2/\kappa$ is small enough so that there are two real modes. On the r.h.s.,  $\mu^2/\kappa$ is large enough so that there is only one real solution and two complex ones.}
\label{plots2}
\end{figure}

To finish this section,  now we consider the full theory \eqref{fulltheory} where  $\phi$ is coupled to a regular variable $x$. 
It can be easily shown that the vibrating modes (where $x$ and $\phi$ are proportional to $\exp (i\omega t)$) have frequencies $\omega$ which satisfy the equation,
\bea
\kappa (\omega^2-\omega_0^2) (\omega^2-\omega_1^2) (\omega^2-\omega_2^2) =  \mu^2 \omega^6 \, .
\eea
We start with the case where $\omega_1 \neq \omega_2$. 
It is easy to see that this equation admits one, two or three solutions for $\omega^2$ according to the value of $\mu^2/\kappa$
as illustrated in Fig. (\ref{plots2}). 
The case with two solutions is singular and corresponds to a fine tuning of the parameters,
thus we will ignore it here. The case with only one solution leads necessarily to a classical instability because two  modes $\omega^2$
become complex. When there are three solutions, we can decouple the three modes, compute the Hamiltonian
as a sum of three independent components (one for each degree of freedom)  and one of these components can be shown to 
have the wrong sign,
exactly as in the case where $\mu=0$. 

One of the easiest way to see this is indeed the case consists in reformulating the Lagrangian
\eqref{fulltheory} with \eqref{omega0} and \eqref{alphabeta}. First, we introduce the new variable $X \equiv x + \mu \dot\phi /\kappa$ 
and we consider the equivalent Lagrangian,
\bea
 - \dot \phi \dot\psi - \frac{\kappa}{2(\kappa-\mu^2)} \psi^2 + \frac{1}{2} \left[ -(\omega_1^2+\omega_2^2) \dot\phi^2 + \omega_1^2\omega_2^2 \phi^2 + \kappa \dot X^2 - \kappa \omega_0^2 (X- \frac{\mu}{\kappa} \dot\phi)^2\right] \, .
\eea
Then, we extract the kinetic part of the Lagrangian (which contains
terms quadratic in the velocities) that can we written in a matrix form as follows,
\bea
L_{\rm kin} \,  = \, \frac{1}{2}\left( \dot \phi , \dot \psi, \dot X\right) 
\left(
\begin{array}{lll}
2 \alpha & -1 & 0 \\
-1 & 0 & 0 \\
0 & 0 & \kappa
\end{array}
\right)
\left(
\begin{array}{c}
\dot \phi \\
\dot \psi \\
\dot X
\end{array}
\right) \, , \quad 2 \alpha \equiv { \mu^2\omega_0^2}/{\kappa} -(\omega_1^2+\omega_2^2) \, .
\eea
where the three dimensional kinetic matrix can be shown immediately to have two positive $\lambda_1$ and $\lambda_2$, and one negative eigenvalues $\lambda_3$ given by,
\bea
\lambda_1=\kappa \, , \qquad \lambda_2=  \alpha+\sqrt{1+\alpha^2}  \, , \qquad
\lambda_3 =  \alpha- \sqrt{1+\alpha^2}  \, .
\eea
 The negative eigenvalues $\lambda_3$ is associated to the kinetic energy of the Ostrogradsky ghost.

The case where $\omega_1=\omega_2$ is even more problematic. Indeed, as it can be seen in Fig.  \eqref{plots2}, there
are always two complex solutions for $\omega^2$ which lead obviously to classical instabilities. Hence, this is an interesting example where
the coupling to a regular variable triggers the Ostrogradsky instability already at the classical level. In the quantum theory, this instability turns into a problem of unitarity. 

As a conclusion, whatever the values of $\omega_1$ and $\omega_2$ are, one cannot escape the instability!

\section{Ostrogradsky theorem for covariant theories}

\label{sec:covtheories}
In this section, we discuss the fate of the Ostrogradsky instability in higher derivative covariant theories. Covariant theories
are invariant under diffeomorphisms and therefore have a vanishing Hamiltonian (if we neglect boundary terms). Therefore, it is interesting
to understand how the instability shows up in this context because, as  the Hamiltonian is vanishing, it cannot be by definition unbounded. Hence, the 
instability cannot be directly and naively linked to the unboundness  of the energy spectrum as it was the case in  previous sections. Then, one can 
legitimately ask the question whether it is problematic to have an Ostrogradsky ghost in covariant theories, and then in scalar-tensor theories.

Addressing this issue in full generality for scalar-tensor theories  goes much beyond the scope of this paper because this would
need a deep understanding of their quantization  which is far from being a trivial task. Instead, we introduce and study  some toy models still in the context of classical mechanics. More precisely, we propose two different ways to ``covariantize'' the classical mechanics models we have studied in previous sections in order to construct theories which are invariant under time reparametrization. Then, we study the fate of Ostrogradsky ghosts in these two frameworks.

\subsection{Theories with a parametrized time}
Let us first discuss the case of a system with a parametrized time. Considering a generic Lagrangian ${L}_0(\dot \phi,\phi)$, for the dynamical variable $\phi(t)$, we can make the system invariant under time reparametrization by introducing an artificial new time variable $\tau(t)$ from which we construct the new Lagrangian,
\begin{eqnarray}
 {L}(\dot\phi,\phi,\dot\tau)\; = \; \dot \tau \, {L}_0 ({\dot \phi}/{\dot \tau} ,\phi)\,.
\end{eqnarray} 
This theory is now clearly invariant under time reparametrization. To perform its canonical analysis, we replace $L$ by the equivalent
Lagrangian,
\begin{eqnarray}
 {L}_{eq}(\dot\phi,\phi,\dot\tau, \pi_\tau,N)\; = \;N \, {L}_0 ( {\dot \phi}/N,\phi ) + \pi_\tau (\dot \tau - N)\,,
\end{eqnarray} 
where $\pi_\tau$ is the momentum conjugate to $\tau$ and $N$ can be shown to be a Lagrange multiplier. 
A straightforward analysis shows that its Hamiltonian is of the form $N H(\phi,\pi;\tau,\pi_\tau)$, and then $H$ is 
a first class constraint given by
\begin{align}
\label{eq:Hamiltonian_parametrized_time}
H(\phi,\pi;\tau,\pi_\tau)\; =\; H_0(\phi,\pi) + \pi_\tau \approx 0 \, ,
\end{align}
where $H_0(\phi,\pi)$ is the Hamiltonian associated to $L_0$ and the Poisson bracket $\{\phi,\pi\}=1$. Of course, $H$ is the generator of time reparametrizations
and the gauge choice $C= \tau -t \approx 0$ allows us to recover the original system $L_0$. As a consequence, the covariant theory associated to $L$ is equivalent to the original one associated to $L_0$.

We can easily adapt this construction to higher-order Lagrangians $L_0(\ddot\phi,\dot\phi,\phi)$ and then introduce
covariant higher-order theories given by,
\begin{eqnarray}
 {L}(\ddot\phi,\dot\phi,\phi,\dot\tau)\; = \; \dot \tau \, {L}_0 \left[\frac{1}{\dot\tau}\frac{d}{dt}(\dot\phi/\dot\tau),{\dot \phi}/{\dot \tau} ,\phi\right]\,.
\end{eqnarray} 
An analysis similar to the previous one shows that its Hamiltonian vanishes as well and it generates time reparametrizations. Furthermore,
the gauge choice $C= \tau -t \approx 0$ reduces the covariant theory to the original one. Hence, there is no reason to expect that
the instabilities present in the original theory will disappear in the covariant one. Indeed, if there is an Ostrogradsky ghost in the theory associated to $L_0$, the covariant Lagrangian $L$ will also propagate the problematic Ostrogradsky ghost.

In order to illustrate this situation, let us discuss the ``covariant'' Pais-Uhlenbeck oscillator whose Lagrangian is given by
\begin{align}
{L}\;=\; \frac{1}{2\dot \tau} \left[ \frac{d }{d t} \left( {\dot \phi}/{\dot \tau}\right) \right]^2 - \frac{1}{2} (\omega_1^2 +\omega_2^2) \frac{\dot \phi^2}{\dot\tau} + \frac{1}{2} \omega_1^2\omega_2^2 \dot\tau \phi^2 \,.
\end{align}
If one introduces the new variable $q= \dot \phi/\dot \tau$ by virtue of a Lagrange multiplier, one can make the canonical analysis and
obtain the first class Hamiltonian constraint which now contains two potentially dangerous terms linear in the momenta,
\begin{align}
\label{eq:Time_invariance_Pais_Uhlenbeck}
H \;=\; \frac{1}{2} p^2 + \frac{1}{2} ( \omega_1^2 +\omega_2^2 ) q^2 - \frac{1}{2} \omega_1^2 \omega_2^2 \phi^2 + \pi q + \pi_\tau \approx 0\, ,
\end{align}
with the following three pairs of conjugate variables,
\begin{eqnarray}
\{\phi,\pi\} = 1 \, , \qquad
\{q,p\}=1 \, , \qquad
\{\tau,\pi_\tau\} = 1 \, .
\end{eqnarray}

For theories with unequal frequencies, $\omega_1 \neq \omega_2 $, we can diagonalize the system and after a redefinition of variables
similar to \eqref{changex1x2}, the Hamiltonian constraint can be reformulated as follows,
\begin{align}
\frac{1}{2} \left( \frac{p_1^2}{m} +  m \omega_1^2 x_1^2 \right) - \frac{1}{2} \left( \frac{p_2^2}{m} + m \omega_2^2 x_2^2 \right) + \pi_\tau \approx 0\,,
\end{align}
where $m$ is defined as in the previous sections by $m \equiv (\omega_1^2-\omega_2^2)^{-1}$, assuming that $\omega_1>\omega_2$, and
$\{x_1,p_1\}=1=\{x_2,p_2\}$.
At the quantum level, we see that, without going into details, quantum states are solutions of the standard Schr\"odinger equation (with time $\tau$) for two harmonic oscillators, one with positive, and one with negative energies. Therefore, the system is still unstable. This is not surprising since we know that the covariant system can be shown to be equivalent to the original Pais-Uhlenbeck oscillator with a gauge fixing. 

For theories with equal frequencies, $\omega_1=\omega_2 =\omega$, the conclusion is the same. In that case, the 
Hamiltonian constraint can be expressed as follows
\begin{align}
 \frac{1}{2} \left( p_1^2 +p_2^2  \right) +\omega L + \pi_\tau \;\approx\; 0\,,
\end{align}
where we used the notations  \eqref{classicalPoi} and \eqref{HamilL}.
Using the same argumentation as before, it is straightforward to conclude that the system is still unstable.

Hence, even though the Hamiltonian is vanishing (and thus not unbounded), the Ostrogradsky instability appears in this type of 
covariant higher order theories.

\subsection{Covariantization with a lapse function}
\label{DHOST}

There is another way to make covariant higher order theories in the context of classical mechanics. One  adds to the theory a new dynamical variable in the form of a  lapse function $N(t)$, and replaces the original
Lagrangian $L_0(\ddot \phi,\dot\phi,\phi)$ by
\begin{eqnarray}
\label{covlag}
L(\ddot\phi,\dot\phi,\phi,N) = N \, L_0 \left[ \frac{1}{N} \frac{d}{dt}\left( {\dot \phi}/{N}\right) ,  {\dot \phi}/{N}, \phi \right] \, .
\end{eqnarray}
The generalization to several variables is immediate.
As in the previous subsection, the Lagrangian is clearly invariant under time reparametrization and its Hamiltonian vanishes
on shell.  Interestingly,  Higher-Order Scalar-Tensor theories reduce to this type of theories when ones limits their study to cosmology where all fields (geometry and scalar field) are functions of time only. For this reason,
studying the classical and quantum stability of Lagrangians \eqref{covlag} is particularly instructive.

Note, that this covariantization is not equivalent to the previous one. While fixing $\tau=t$ in the previous section is a viable gauge choice leading to the original theory, there is no gauge choice which would transform \eqref{covlag} into the non-covariant theory. Indeed, fixing $N=1$ directly in the Lagrangian is not a proper gauge choice since it is an incomplete gauge fixing which does not uniquely define the gauge transformation \cite{Motohashi:2016prk}.  Instead, as we will later see, the presence of the lapse function leads to two first class constraints as in general relativity, the Hamiltonian constraint ${H_0}$ due to the time reparametrization invariance and the momentum of the lapse function $\pi_N$. Consequently, it requires two gauge fixing conditions while the condition $N=1$ only fixes the first class constraint $\pi_N$. Another more direct heuristic argument to realize that $N=1$ is an incomplete gauge fixing is that in order to fix the time reparametrization invariance it is necessary that the gauge fixing condition explicitly depends on time as in the previous covariantization $\tau=t$ which is not the case for $N=1$. Last, note that due to the two first class constraint the covariant theory has one less degree of freedom than the non-covariant one. 

\medskip

To illustrate the construction, let us start with a simple harmonic oscillator $\phi(t)$, of mass $m$ and frequency $\omega$,
whose covariant Lagrangian is given by,
\begin{align}
{L}(\dot\phi,\phi,N)\; =\; \frac{1}{2 N}  \dot \phi^2 - \frac{1}{2} N \omega^2 \phi ^2\,.
\end{align}
The lapse function $N$ is a Lagrange multiplier and the Hamiltonian $H$ reads,
\begin{align}
H \;=\;  N H_0 \; \equiv \; N \left( \frac{\pi^2}{2} + \frac{1}{2} \omega^2 \phi^2 \right)\;\approx \; 0\,, 
\end{align}
with  $\{\phi,\pi\}=1$. As $H$ is a first class constraints, the system has no degrees of freedom. Further, the Hamiltonian constraint has no valid solution a part from the trivial one, $\phi=0=\pi$. We can generalize it to a sum of two harmonic oscillators yielding one degree of freedom. However, the Hamiltonian constraint has still no valid solutions since $H_0$ is positive definite, which yields an empty Hilbert space at the quantum level. 
\medskip

Let us now consider the covariant Pais-Uhlenbeck oscillator,
\begin{align}
{L}(\ddot \phi, \dot \phi, \phi, \dot N , N) \;=\; \frac{1}{2N}  \frac{d }{d t} \left( {\dot \phi}/{N}\right)^2 - \frac{1}{2N} (\omega_1^2 +\omega_2^2) {\dot \phi^2} + \frac{N}{2} \omega_1^2\omega_2^2  \phi^2\, .
\end{align}
The Hamiltonian analysis of this theory is similar to the analysis of the non-covariant theory (where $N=1$) and we arrive at the same
result with, in addition, the constraint that the energy vanishes which removes one of the two modes.

The quantization is, therefore, straightforward. In the case where $\omega_1 \neq \omega_2$, the quantum states are the same as in section \eqref{quantumPU}, they are labelled by $\vert n_1,n_2 \rangle$ where $n_1$ and $n_2$ are integers, with in addition,  the condition
\bea
\hat{H} \vert n_1,n_2 \rangle = 0  \quad \Longrightarrow \quad \omega_1(2n_1 + 1) = \omega_2 (2 n_2 + 1) \, .
\eea
It is easy to show that this last equation has no solutions except if the two frequencies are equal, otherwise the space of physical
quantum states is empty and then the  quantum theory is ill-defined.

The only hope to have a well-defined model is to consider the degenerate Pais-Uhlenbeck action ($\omega_1=\omega_2=\omega$)
in which case the Hamiltonian is \eqref{HamilL}. Hence, the quantum states are given by \eqref{quantumstates} with $E=0$,
Then, the space of states becomes now discrete (as states are labelled by an integer $\ell$) but they are not normalizable, and  they cannot be interpreted in terms of wave functions. As a consequence, in that case also, the quantum theory is ill-defined.

\medskip

So far, for none our discussed models the quantum theory was well-defined. However, this is not necessarily the case. In the appendix,  we construct a specific toy model where the quantum theory is  well-defined despite the presence of higher time derivatives. Indeed, the additional degree of freedom, introduced by the  higher derivative terms, behaves just as a standard harmonic oscillator with positive energy. 
Let us elaborate it in more detail. The Lagrangian \eqref{covlag} can be equivalently expressed as 
\begin{align}
L_{eq}(\dot q, q,\dot \phi , \phi, \pi,N)\; = \; N L_0 ( \dot q , q , \phi) + \pi ( \dot \phi - N q)\,,
\end{align}
where $\pi$ is the momentum conjugate to $\phi$. A straightforward analysis shows that its Hamiltonian is of the form $N H(\phi,\pi;q,p)$, and $H$ is a first class constraint given by
\begin{align}
 H (\phi,\pi;q,p) \;=\; H_0 ( \phi; q,p) + q \pi\,, 
\end{align}
where $H_0$ is the Hamiltonian associated to $L_0$ with $\{q,p\}=1=\{\phi,\pi\}$. We can note that the fundamental structure is similar to \eqref{eq:Hamiltonian_parametrized_time}, in particular, if the original model does not explicitly depend on $\phi$. 
Comparing to the analysis of models with a parametrized time, it is immediate that one can construct models with a well-defined quantum theory. The term linear in the momentum of $\phi$ does not necessarily imply an instability but instead leads to a standard Schr\"odinger equation.

These results are not in contrast to the well-known Ostrogradsky theorem, since the covariant models are also degenerate, in the sense that the lapse function is not dynamical and introduces two first class constraints which can remove the Ostrogradsky ghost. 
Note that, in the framework of DHOST commonly a different nomenclature is used. In this case, the models are called degenerate if the kinetic matrix without the lapse function and the shift vector is non-invertible which leads to additional constraints besides the standard eight first class constraints from the diffeomorphism invariance \cite{Langlois:2015cwa}. 

\section{Discussion}
\label{conclusion}

The Ostrogradsky ghost has been at the center of many concerns in recent years, more particularly, in the context of general relativity and
theoretical cosmology where higher order theories are often used for several physical reasons.  For a long time, it was commonly thought that the presence of non-trivial higher derivatives 
in the Lagrangian  would necessarily lead to lethal pathologies. As we recalled in this paper, this is indeed true for higher derivative theories
of a sole variable in classical mechanics, as stated by the famous Ostrogradsky theorem. In that case, higher derivatives theories have necessarily an unbounded Hamiltonian which obviously leads to fundamental instabilities. However, this is no more necessarily true when
one considers theories with multi-variables in classical mechanics, and even less true  for field theories. 

In this paper, we mainly focused on the case of higher order theories with multiple variables in classical mechanics. We recalled that, in spite of the presence
of non trivial higher derivatives in the Lagrangian, and even if the Euler-Lagrange equations are higher order, the theory may be totally well defined with no propagation of any Ostrogradsky ghosts when the Lagrangian is degenerate. More precisely, the degeneracy is a necessary condition to evade Ostrogradsky instabilities for higher order theories. Only a further study in a case by case basis would enable us to
conclude that the degeneracy is also a sufficient condition because one has to be sure that the degeneracy condition kills the Ostrogradsky
ghost and not a well behaved degree of freedom. We illustrated all these aspects in this paper and we also reported on the quantum 
pathologies of the Ostrogradsky ghost for free higher order theories (consisting in the Pais-Uhlenbeck model and its generalizations).  Finally, we discussed new aspects concerning the case of higher order covariant theories. Due to the time-reparametrization invariance the Hamiltonian is perfectly bounded as it vanishes on shell. The presence of higher derivatives do not necessarily imply an ill-defined quantum behavior. Instead, the two first class constraints can be sufficient to remove the Ostrogradsky ghost and the additional degree of freedom due to the higher derivatives in the Lagrangian is well-defined. Note, that it is not necessary to require additional degeneracy conditions as in the construction of DHOST which we have quickly presented in the introduction. 

%which are particularly interesting because, if non degenerate, they propagate an Ostrogradsky ghost while their Hamiltonian is perfectly bounded as it vanishes on shell. So it was important to understand how the instability shows up in this case, which is what we have presented. 

\medskip

We have not discussed, in this paper, the case of field theories which are physically more interesting. Nonetheless, the results we have presented here in the context of classical mechanics generalize easily to the case of field theories even though there are  
differences compared to classical mechanics. One  important difference is the possibility to construct higher order field theories with only one
scalar field which do not propagate any Ostrogradsky ghosts: these are the famous Galileon theories which have been defined in flat or curved space-time \cite{Nicolis:2008in,Deffayet:2009wt,Deffayet:2009mn} and in any space-time dimension $D+1$ (here we restrict our discussion to $D=3$ for simplicity). They have been designed in such a way that their corresponding equation of motion is second
order. But the deep reason why such theories do not have Ostrogradsky instabilities lies in the fact that, if one reformulates the Galileon 
action using a 3+1 decomposition of the space-time, one immediately sees that the Lagrangian does not contain any higher order time derivatives of the scalar field. Higher order derivatives in the Lagrangian are always ``cross'' partial derivatives of the form $\partial_\mu \partial_\nu \phi$ with different indices, i.e. $\mu \neq \nu$. In that sense, there is no higher-order time derivatives and one understands why there is no
Ostrogradsky ghost. 

There are many generalizations of Galileon theories. One can construct multi-Galileon models for instance or make the background dynamical which leads to the celebrated Horndeski theories and then DHOST theories. DHOST theories can be viewed as  generalizations of multi-variables degenerate higher order theories in classical mechanics where the two fields (a scalar and a tensor) interact in such a way that the Ostrogradsky ghost is never excited. They encompass Horndeski theories, beyond Horndeski and also arbitrary disformal transformations of Horndeski theories, but their Euler-Lagrange equations are no longer second order in general. DHOST theories have been used intensively in the context of cosmology and astrophysics (see \cite{Langlois:2017mxy,Crisostomi:2018bsp,Crisostomi:2017pjs,Boumaza:2020klg,Babichev:2016jom} for instance and the reviews \cite{Langlois:2018dxi,Kobayashi:2019hrl}). Most of the literature has recently concentrated on DHOST theories where the speed of gravitational waves coincides with that of light \cite{TheLIGOScientific:2017qsa,Creminelli:2017sry,Sakstein:2017xjx,Ezquiaga:2017ekz,Crisostomi:2017lbg,Langlois:2017dyl,Kobayashi:2018xvr}. We would like to mention the possibility to extend the class
of DHOST theories to the so-called U-degenerate theories \cite{DeFelice:2018ewo} in which the unitary gauge (where the scalar field coincides with the time variable) leads to additional constraints removing the Ostrogradsky ghost. Even if these constraint are not present in their fully covariant version, it was argued that
the extra mode is, in fact, a so-called shadowy mode which does not propagate, but instead its ``dynamics'' is governed by an elliptic partial differential equation. Finally, let us quickly mention the results of \cite{Crisostomi:2017ugk} where it was shown that
it is impossible to combine the different components of the metric to build new (invariant under diffeomorphisms) 
degenerate theories for the metric only. 

\medskip

To conclude, let us emphasize again the main message of the paper. Higher order Lagrangians and/or higher order equations of motion
do not necessarily imply the existence of an Ostrogradsky ghost. However, the degeneracy is a necessary condition to eliminate the
Ostrogradsky instability from a higher order theory.

\section*{Acknowledgements}
We would like to thank 
 Marco Crisostomi and David Langlois for  their contributions to the topic discussed in this paper, for their suggestions and
for their participations at the early stages of the project. 
We also want to warmly acknowledge Hugo Roussille for valuable and very interesting discussions, and also for helping resolving 
Eq. \eqref{EqLcase1}. A.G wishes to thank APC for hospitality during the development of this project. K.N. wants to thank the Laboratoire de Physique Th\'eorique at the ENS and especially Costas Bachas,  for hospitality during the development of this project. 

\appendix
\section*{Appendix: Covariant toy model}
\label{app:Toy_model_stable}
So far the covariant systems we have considered have no valid quantum theories. Here, we propose a model 
that works well. To introduce it, let us consider now a more generic case,
\begin{align}
\mathcal{L}\;=\; \frac{A(q,x)}{2N} \dot q^2 + \frac{K(q,x)}{2N} \dot x^2 - N V(q,x) + \pi ( \dot \phi - N q)\,.
\end{align}
As before, we obtain two first class constraints,
\begin{align}
 H_0 \; =\;  \left( \frac{1}{2 A} p^2 + \frac{1}{2 K} p_x^2 +  V + q \pi \right) \approx 0\,, \qquad
\pi_N\; \approx\;  0\,,
\end{align}
with the Poisson structure,
\bea
\{q,p\} \, = \, \{x,p_x\} \, = \, \{\phi,\pi\} \, = \, 1\,.
\eea
We obtain one term linear in the momentum of $\pi$ signaling normally an instability. Let us focus on the very specific sub-case with,
\begin{align}
A\;=\; \frac{N}{\dot \phi}\,, \qquad K\;=\;\frac{N}{\dot \phi}\,, \qquad V \;=\; \frac{\dot\phi}{2 N} \left( x^2 + \frac{\dot \phi^2}{N^2}\right)\,.
\end{align}
In this case the two first class constraints can be expressed as, 
\begin{align}
H^\prime \;=\;  q^{-1} H_0 \;=\; \frac{1}{2} p^2 + \frac{1}{2}  p_x^2 + \frac{1}{2}  x^2 + \frac{1}{2} q^2 + \pi \;\approx\; 0\,, \qquad
\pi_N \;\approx\;  0\,.
\end{align}
The second one can be solved trivially, while from the first class Hamiltonian we obtain the standard Schr\"odinger equation for the sum of two harmonic oscillators with positive energy. The system is totally well defined. Note that this is not in contrast to the Ostrogradsky ghost theorem since our original model is degenerate due to the fact that the lapse function $N$ is non-dynamical and its momentum vanishes.

It is interesting to see what happens if we turn off the higher derivative term by setting \\ $A(q,x)=0$. 
In this case we obtain for the Lagrangian,
\begin{align}
\mathcal{L} \;=\; \frac{\dot x^2}{2 \dot \phi} - \frac{\dot \phi}{2} \left( x^2 + \frac{\dot \phi^2}{N^2} \right)\,.
\end{align}
The two first class constraints can be expressed as $\pi_N\approx 0$ and
\begin{align}
H \;=\;  N (2 \pi + p_x^2 + x^2)^{3/2} \;\approx\; 0\,.
\end{align}
Before analyzing it further, let us fix the frame by introducing the constraint $C=N-1\approx 0$, which forms together with the constraint $\pi_N\approx 0$ a set of two second class constraints. Setting strongly $\pi_N=0$ and $N=1$ the extended Hamiltonian is given by
\begin{align}
H_T \;=\; u_H (2 \pi + p_x^2 + x^2)^{3/2} \, .
\end{align}
The classical equations of motion read
\begin{align}
& \dot p_x \;=\; \{p_x , \tilde H_T \} \;=\; - 3 u_H x \sqrt{2 \pi + p_x^2 + x^2} \, , \qquad & \dot x \;=\; \{ x, \tilde H_T \} \;=\; 3 u_H p_x \sqrt{2 \pi + p_x^2 + x^2}\,, \\
& \dot \pi \;=\; \{ \pi , \tilde H_T \} \;=\; 0\, , \qquad & \dot \phi \;=\; \{ \phi, \tilde H_T \} \;=\; 3 u_H \sqrt{2 \pi + p_x^2 + x^2}\,. 
\end{align}
Using the chain rule we obtain
\begin{align}
&  \frac{\dot p_x}{\dot \phi} \;=\; \frac{d p_x(\phi(t))}{d \phi} \;=\; - x(\phi(t)) \,, &   \frac{\dot x}{\dot \phi} \;=\; \frac{d x(\phi(t))}{\mathrm{d} \phi} \;=\; p_x(\phi(t))\,,
\end{align}
which are the standard differential equations for a time invariant harmonic oscillator.

We can conclude that turning on the higher time derivative term has added one standard harmonic oscillator with positive energy. Therefore, the higher time derivative terms in such kind of systems does not signalize necessary an instability. Note, that it is related to the fact that there is only one potential dangerous term linear in the momentum which can be removed by the Hamiltonian constraint.

\bibliographystyle{utphys}
\bibliography{Ostro_biblio}
\end{document}